\documentclass[conference]{IEEEtran}
\IEEEoverridecommandlockouts

\usepackage{cite}
\usepackage{amsmath,amssymb,amsfonts}
\usepackage{algorithmic}
\usepackage{graphicx}
\usepackage{textcomp}
\usepackage{xcolor}
\usepackage{float}
\usepackage{subcaption} 
\usepackage[flushleft]{threeparttable} 
\usepackage{booktabs}    
\usepackage{tabularx} 
\usepackage{makecell} 
\usepackage[inline]{enumitem} 
\usepackage{booktabs}
\usepackage{longtable}        
\usepackage{threeparttablex}  
\usepackage{arydshln} 
\usepackage[hidelinks]{hyperref} 
\def\BibTeX{{\rm B\kern-.05em{\sc i\kern-.025em b}\kern-.08em
    T\kern-.1667em\lower.7ex\hbox{E}\kern-.125emX}}
\begin{document}

\title{VissimRL: A Multi-Agent Reinforcement Learning Framework for Traffic Signal Control \\Based on Vissim}

\author{
    Hsiao-Chuan Chang\textsuperscript{1*},
    Sheng-You Huang\textsuperscript{1*},
    Yen-Chi Chen\textsuperscript{2}, and
    I-Chen Wu\textsuperscript{1,2,3}

    \thanks{This work was supported in part by the National Science and Technology Council (NSTC) of Taiwan under Grant, NSTC 114-2221-E-A49-005, NSTC 114-2221-E-A49-006.}
    
    \thanks{\textsuperscript{1}Department of Computer Science, National Yang Ming Chiao Tung University, Hsinchu, Taiwan.}
    \thanks{\textsuperscript{2}Research Center for Information Technology Innovation, Academia Sinica, Taipei, Taiwan.}
    \thanks{\textsuperscript{3}Correspondence (e-mail: icwu@cs.nycu.edu.tw)}
    \thanks{\textsuperscript{*}Equal contribution.}
}

\maketitle
\begin{abstract}
Traffic congestion remains a major challenge for urban transportation, leading to significant economic and environmental impacts. 
Traffic Signal Control (TSC) is one of the key measures to mitigate congestion, and recent studies have increasingly applied Reinforcement Learning (RL) for its adaptive capabilities. 
With respect to SUMO and CityFlow, the simulator Vissim offers high-fidelity driver behavior modeling and wide industrial adoption but remains underutilized in RL research due to its complex interface and lack of standardized frameworks. 
To address this gap, this paper proposes VissimRL, a modular RL framework for TSC that encapsulates Vissim's COM interface through a high-level Python API, offering standardized environments for both single- and multi-agent training. 
Experiments show that VissimRL significantly reduces development effort while maintaining runtime efficiency, and supports consistent improvements in traffic performance during training, as well as emergent coordination in multi-agent control.
Overall, VissimRL demonstrates the feasibility of applying RL in high-fidelity simulations and serves as a bridge between academic research and practical applications in intelligent traffic signal control.
\end{abstract}

\begin{IEEEkeywords}
Reinforcement Learning, Traffic Signal Control, Vissim, Multi-Agent Coordination, Modular framework.
\end{IEEEkeywords}

\section{Introduction}
Traffic congestion remains a critical challenge for modern cities, causing substantial economic and environmental impacts.
The European Union estimated that congestion resulted in annual losses of about €270 billion \cite{europeancourtofauditorsSustainableUrbanMobility2020a}, while in the United States, congestion in 2022 led to 8.5 billion hours of delay, 3.3 billion gallons of excess fuel consumption, and approximately \$224 billion in losses \cite{schrank2023UrbanMobility2024}. 
In London, congestion produced 2.2 million metric tons of CO\textsubscript{2} in 2021, accounting for roughly 15\% of the city's total emissions \cite{beedhamTrueEnvironmentalCost2022}, underscoring the urgency of effective traffic management. 
Traffic Signal Control (TSC) offers a cost-effective and scalable solution, and Reinforcement Learning (RL) has shown great potential by enabling autonomous policy learning without explicit traffic models \cite{zhaoSurveyDeepReinforcement2024}. 
Since the introduction of the Deep Q-Network (DQN) in 2013 \cite{mnihPlayingAtariDeep2013}, Deep Reinforcement Learning (DRL) has become an important research direction, with proven effectiveness in both single-intersection \cite{shashiStudyDeepReinforcement2021,oroojlooyAttendLightUniversalAttentionBased2020} and multi-intersection control \cite{weiPressLightLearningMax2019,weiCoLightLearningNetworklevel2019}. 
Nevertheless, real-world deployment of reinforcement learning for traffic signal control (RL-TSC) requires highly realistic and reproducible simulation environments to support large-scale training and evaluation.

Traffic simulation tools are essential for developing and evaluating TSC methods, as real-world systems cannot be frequently adjusted or tested. 
Simulations provide a safer and more feasible alternative, operating dozens to over a hundred times faster than reality \cite{schraderExtensionValidationNEMAStyle2022}. 
Commonly used platforms include SUMO \cite{krajzewiczSUMOSimulationUrban2002}, CityFlow \cite{zhangCityFlowMultiAgentReinforcement2019}, and Vissim \cite{fellendorfMicroscopicTrafficFlow2010}. 
SUMO is open-source and widely used in academia; CityFlow is designed for RL experiments with high efficiency and scalability; and Vissim, as a commercial-grade software, provides highly realistic modeling of driver behaviors and complex traffic scenarios \cite{baggioComparisonSimulatorsSUMO2023c}. 
Owing to its superior accuracy, Vissim has been adopted in official guidelines by various U.S. state Departments of Transportation \cite{schilperoortProtocolVISSIMSimulation2014,hunterVISSIMSimulationGuidance2021,hillDevelopmentMichiganSpecificVISSIM2020,iowadepartmentoftransportationIowaDOTMicrosimulation2017} and also recommended by Transport for London \cite{beestonTrafficModellingGuidelines2021}.

Research on RL-TSC has mostly employed open-source simulators such as SUMO and CityFlow \cite{michailidisTrafficSignalControl2025}, given their accessibility and scalability. 
However, their rule-based driver behavior models often oversimplify traffic dynamics \cite{daCityFlowEREfficientRealistic2024}, and their relevance to industry and policy is limited. 
By contrast, the simulator Vissim offers industry-grade realism but is rarely used in RL research due to its complex Component Object Model (COM) programming interface \cite{whiteCOMObjectsInterfaces} and the lack of RL frameworks for seamless integration.  

To address these gaps, this paper adopts Vissim as the primary simulation platform and develops an RL-compatible framework that leverages its modeling accuracy while enabling broader applicability in both academic research and practical deployment.
In this context, SUMO-RL \cite{sumorl} has been a widely used open-source RL framework for SUMO, offering standardized interfaces for RL environments compatible with Gymnasium \cite{towersGymnasiumStandardInterface2024} for single-agent and PettingZoo \cite{terryPettingZooGymMultiAgent2021} for multi-agent scenarios, and supporting flexible customization of states and rewards. 
Its user-friendliness and extensibility have made it a foundational platform for RL-TSC research. 

This paper proposes VissimRL, an RL framework designed for TSC in the Vissim environment, which achieves the same user-friendliness and extensibility as SUMO-RL, while delivering higher modeling fidelity suitable for industrial applications.
The framework adopts a flexible modular architecture to support diverse TSC applications, thereby aiming to close the gap between academic research and real-world deployment.

The architecture of VissimRL consists of two major components: (1) the Vissim Wrapper, a Python-based encapsulation of the Component Object Model (COM) interface that provides user-friendly APIs; and (2) the RL environment framework, which implements standardized Gymnasium and PettingZoo interfaces for single- and multi-agent scenarios.
Compared with SUMO-RL, VissimRL introduces modular designs of states, actions, and rewards with standardized inheritance interfaces for easy customization.
In addition, VissimRL extends SUMO-RL by supporting performance metrics from six to eleven and expanding one action type to three action types, as shown in Table \ref{tab:rltsc_framework_comparison}. 
To the best of our knowledge, VissimRL is the first framework specifically designed for Vissim with multi-agent RL, enabling RL in Vissim's high-fidelity traffic simulation environment. 
It seamlessly supports mainstream algorithms, e.g., DQN\cite{mnihPlayingAtariDeep2013}, PPO\cite{schulmanProximalPolicyOptimization2017}, MAPPO\cite{yuSurprisingEffectivenessPPO2022}, without additional glue code, and provides a flexible state–action–reward pipeline, substantially enhancing the applicability of Vissim in RL-TSC research.

This paper is an extended version of the work published in \cite{Chang_VissimRL_IV}. The implementation is publicly available at {\small\texttt{https://github.com/CGI-LAB/VissimRL}}. 

\begin{table}[h]
    \centering
    \caption{Comparison of supported features between SUMO-RL and VissimRL (bold items indicate features additionally implemented in this work).}
    \label{tab:rltsc_framework_comparison}
    \begin{tabularx}{\linewidth}{p{1.5cm} p{3.8cm} X}
    \toprule
    & \textbf{State/Reward metrics} & \textbf{Action types} \\ \midrule
    \textbf{SUMO-RL supported} 
        & Queue length, Phase state, Number of vehicles, Density, Speed, Internal waiting time 
        & Choose next phase \\ \midrule
    \textbf{VissimRL supported} 
        & Queue length, Phase state, Number of vehicles, Density, Speed, Internal waiting time, \textbf{Boundary waiting time, Elapsed time, Delay, Throughput, Travel time} 
        & Choose next phase \newline 
          \textbf{Switch next or not} \newline 
          \textbf{Set phase duration} \\ 
    \bottomrule
    \end{tabularx}
\end{table}

\section{Related Work}

\subsection{Traffic Simulation Software}
SUMO \cite{krajzewiczSUMOSimulationUrban2002}, developed by the German Aerospace Center in 2002, is one of the most widely used microscopic traffic simulators in academia \cite{michailidisTrafficSignalControl2025}. 
As an open-source and cross-platform tool, SUMO provides a Python library (sumolib \cite{dlrandcontributorsSumolibPythonHelper}) for handling network and simulation data, along with a wide range of modular utilities for constructing networks and traffic scenarios.
For integration with RL, the open-source project SUMO-RL \cite{sumorl} provides a standardized RL environment that allows flexible customization of states and rewards.
These features have made SUMO-RL one of the representative and mature RL-TSC platforms.

CityFlow \cite{zhangCityFlowMultiAgentReinforcement2019}, introduced by Zhang et al. in 2019, was specifically designed for large-scale urban traffic scenarios and RL applications. 
It features high-performance simulation with multi-threading support and a native Python API that enables the rapid construction of training environments. 
Unlike SUMO, which benefits from the SUMO-RL framework, to the best of our knowledge, CityFlow currently lacks a widely adopted unified RL framework. 
Nevertheless, its simple API design and RL-friendly characteristics allow researchers to build customized RL environments with relatively low integration costs.

Vissim \cite{fellendorfMicroscopicTrafficFlow2010} is a commercial high-resolution traffic simulation platform capable of modeling multimodal traffic with advanced driver behavior algorithms and 3D visualization. It has been widely adopted by transportation agencies \cite{schilperoortProtocolVISSIMSimulation2014,hunterVISSIMSimulationGuidance2021,hillDevelopmentMichiganSpecificVISSIM2020,iowadepartmentoftransportationIowaDOTMicrosimulation2017,beestonTrafficModellingGuidelines2021} and provides a Component Object Model (COM) interface \cite{whiteCOMObjectsInterfaces} for secondary development. 
Despite its high modeling accuracy, the integration of RL with Vissim remains challenging, primarily due to the absence of standardized and open-source frameworks. Existing studies often rely on tightly coupled implementations, which hinder reproducibility and extensibility across diverse traffic scenarios.


\subsection{Existing RL Framework – SUMO-RL}
SUMO-RL \cite{sumorl} is an RL environment framework specifically designed for TSC. 
It integrates the SUMO traffic simulator with RL algorithms by leveraging SUMO's sumolib module \cite{dlrandcontributorsSumolibPythonHelper} to process network data and enable simulation interaction. 
The framework provides standardized RL interfaces, supporting both the single-agent Gymnasium API \cite{towersGymnasiumStandardInterface2024} and the multi-agent PettingZoo API \cite{terryPettingZooGymMultiAgent2021}, and can be seamlessly combined with mainstream RL libraries such as Stable-Baselines3 \cite{raffinStablebaselines3ReliableReinforcement2021} and RLlib \cite{liangRLlibAbstractionsDistributed2017b}. 
SUMO-RL manages each intersection as an independent traffic signal module and allows users to incorporate custom state and reward functions through configuration files, thereby supporting diverse experimental requirements and enhancing the flexibility of multi-agent system design. 
As a result, SUMO-RL effectively lowers the barrier to applying RL in the TSC domain and has accelerated the development of related algorithms and applications.

\subsection{RL settings for traffic signal control}
In the context of RL, the decision-making process of an agent is primarily defined by three core elements: states, rewards, and actions. 
According to a comprehensive survey of 160 RL-TSC studies conducted by Noaeen et al. (2022) \cite{noaeenReinforcementLearningUrban2022}, the most frequently adopted traffic performance metrics are predominantly utilized in the design of state and reward representations.
Most existing studies on action design focus on controlling traffic signal phases, typically by adjusting phase durations or switching between phases \cite{wangTrafficSignalCycle2024}.
The following subsections describe the typical designs of states, rewards, and actions.

\subsubsection{States}
States represent the observable traffic features perceived by the agent and serve as the basis for decision-making. 
Commonly used state features include:
\begin{itemize}
    \item \textbf{Queue Length}: the number or total length of vehicles moving below a low-speed threshold (e.g., $< 5$ km/h);
    \item \textbf{Phase State}: the current active signal phase at the intersection;
    \item \textbf{Number of Vehicles}: the total number of vehicles per lane, including both queued and moving vehicles;
    \item \textbf{Speed}: the average vehicle speed per lane;
    \item \textbf{Elapsed Time}: the time elapsed since the last phase transition.
\end{itemize}
These features help the agent capture traffic conditions and congestion levels, enabling the design of adaptive control strategies.

\subsubsection{Rewards}
The reward function is the key mechanism that guides the agent's behavior. In RL-TSC, 
commonly used reward indicators include:
\begin{itemize}
    \item \textbf{Queue Length}: a measure of congestion based on the number or length of vehicles traveling below a speed threshold (e.g., $< 5$ km/h);
    \item \textbf{Delay}: the difference between actual travel time and the free-flow travel time, reflecting traffic efficiency loss;
    \item \textbf{Waiting Time}: the total time vehicles spend idle or at low speeds;
    \item \textbf{Throughput}: the number of vehicles passing through the intersection per unit of time;
    \item \textbf{Travel Time}: the aggregate travel time of all vehicles in the road.
\end{itemize}
Most studies employ a weighted combination of multiple indicators to form a composite reward \cite{weiIntelliLightReinforcementLearning2018,vanderpolCoordinatedDeepReinforcement2016,zhengDiagnosingReinforcementLearning2019}, balancing different objectives and more comprehensively reflecting the effectiveness of the control system.

\subsubsection{Actions}
\label{subsubsec:actions}
Actions define how the agent controls traffic signals at each decision step. 
At an intersection, a phase is a set of traffic movements that can proceed simultaneously without conflict, and the essence of signal control lies in managing these phases \cite{wangTrafficSignalCycle2024}.
Using a two-lane, four-phase intersection as an example (Figure \ref{fig:phases}), three representative action types are:
\begin{figure}[H]
    \centering
    \includegraphics[width=0.7\linewidth]{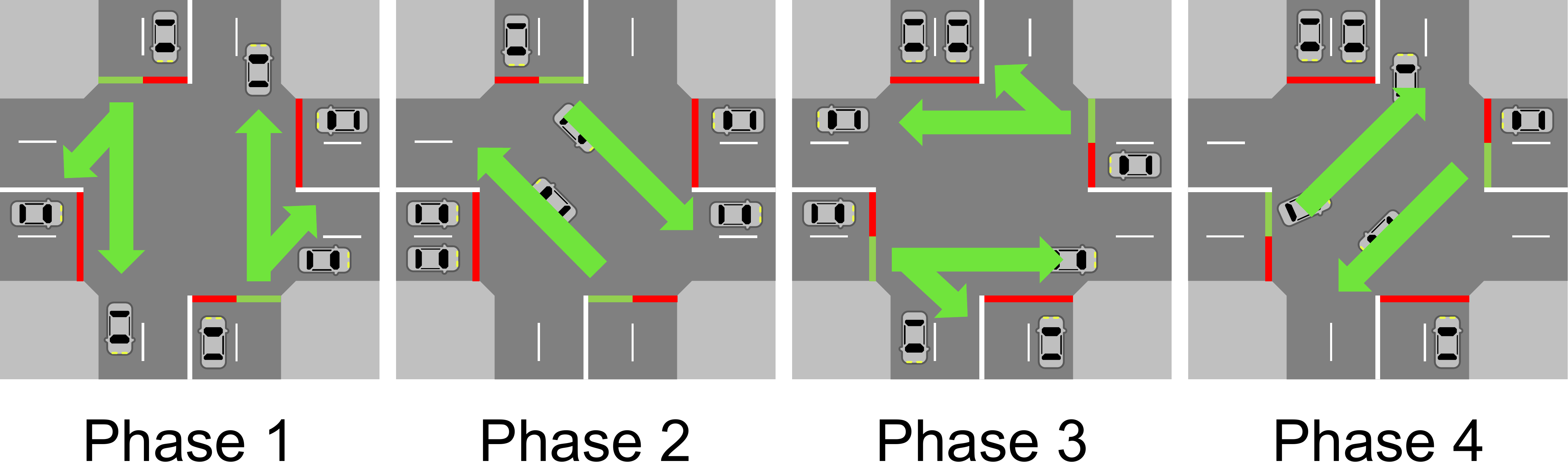}
    \caption{Four typical signal phases at a two-lane intersection.}
    \label{fig:phases}
\end{figure}

\begin{itemize}
    \item \textbf{Choose Next Phase}:\\
    At fixed intervals (e.g., every $n$ seconds), the agent selects the next active phase from all available options, directly determining the switching sequence (Figure \ref{fig:choose_next_phase}). 
    This design allows flexible sequencing based on real-time traffic conditions rather than fixed cycles.
    \begin{figure}[H]
        \centering
        \includegraphics[width=0.95\linewidth]{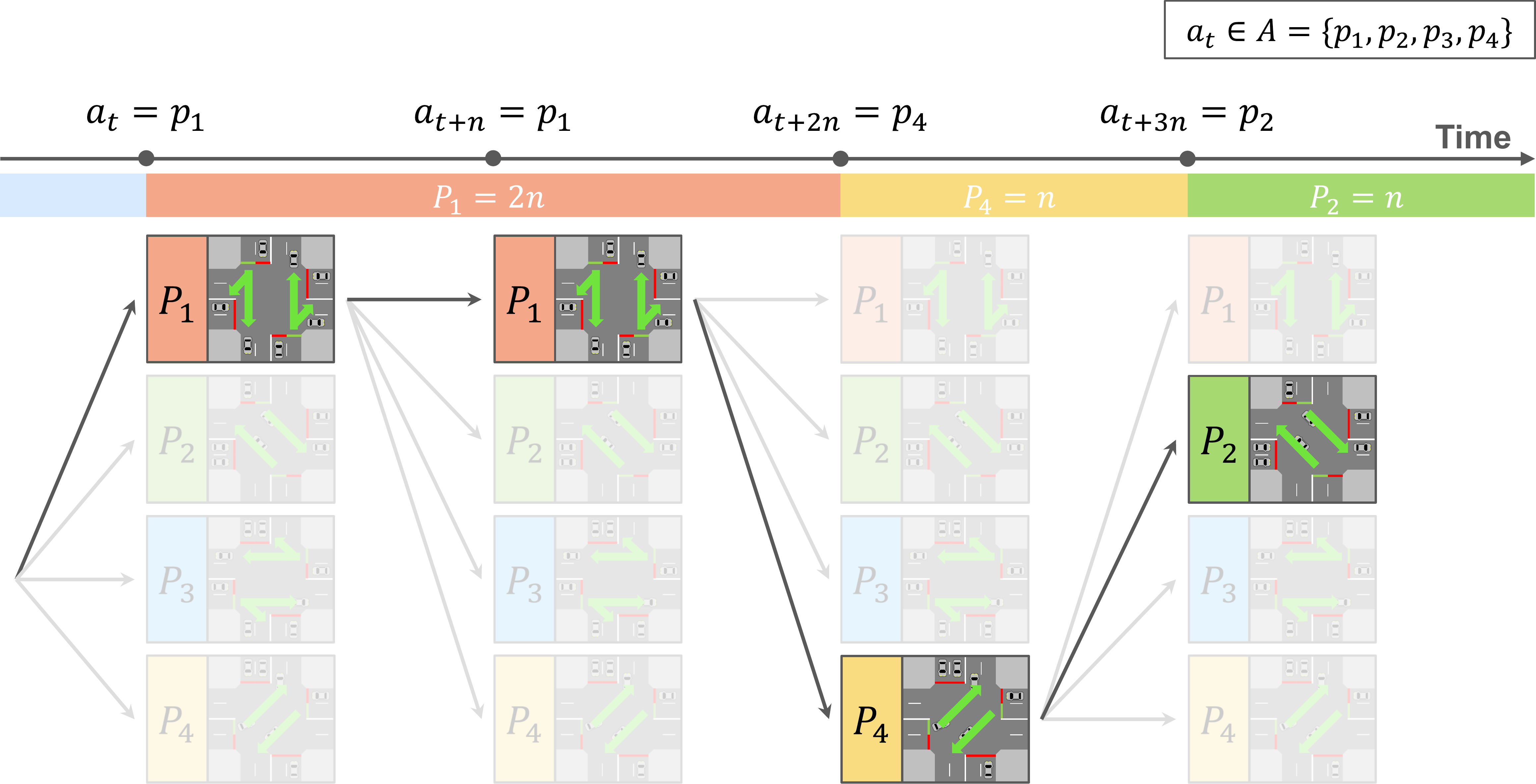}
        \caption{Illustration of the \emph{Choose Next Phase} action design.}
        \label{fig:choose_next_phase}
    \end{figure}
    
    \item \textbf{Switch Next or Not}:\\
    At fixed intervals, the agent decides whether to maintain the current phase or switch to the predefined next phase (Figure \ref{fig:switch_next_or_not}). 
    This design allows temporary extension of the current phase to accommodate heavy flows in a given direction.
    \begin{figure}[H]
        \centering
        \includegraphics[width=0.95\linewidth]{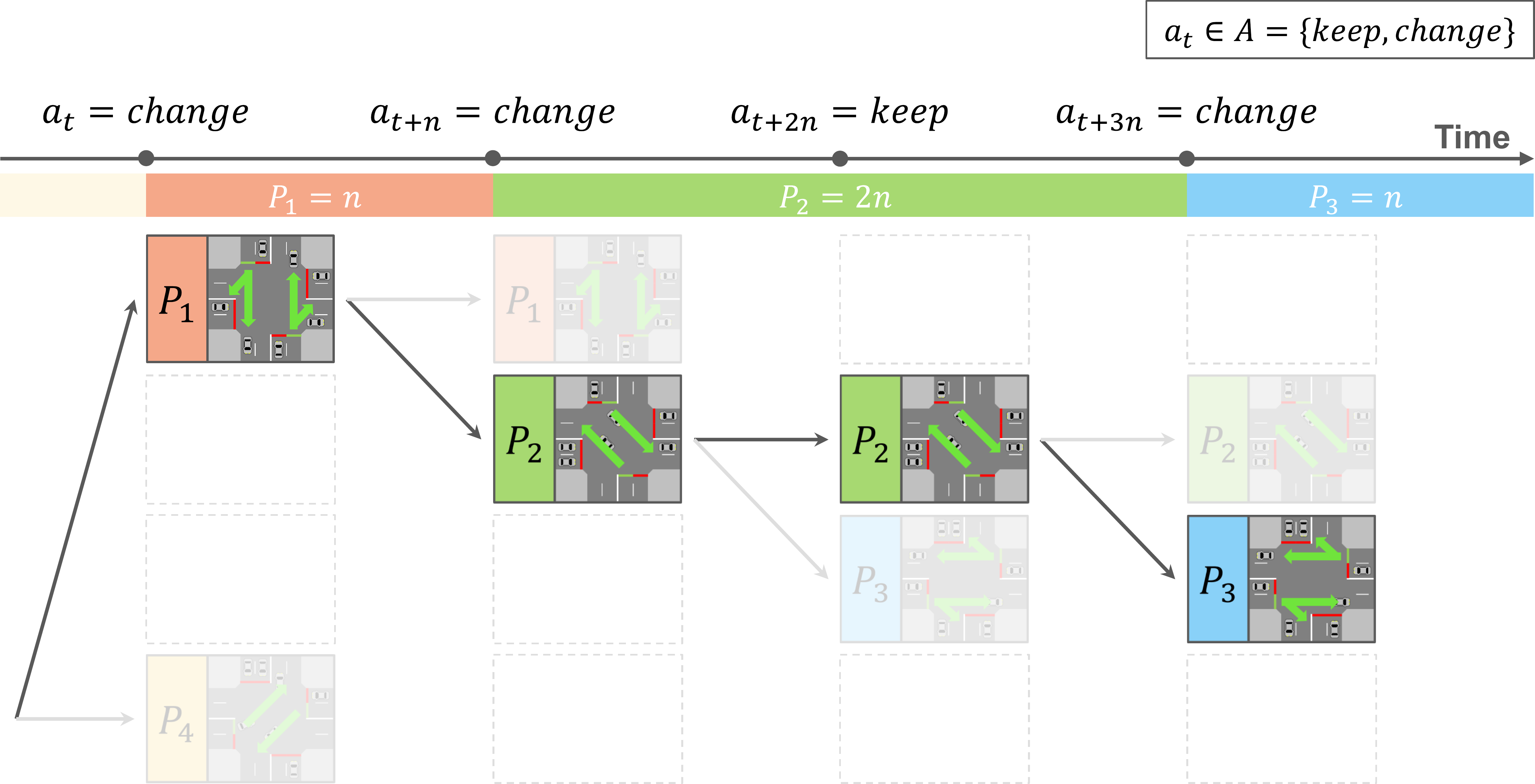}
        \caption{Illustration of the \emph{Switch Next or Not} action design.}
        \label{fig:switch_next_or_not}
    \end{figure}
    
    \item \textbf{Set Phase Duration}:\\
    Before each phase begins, the agent determines the green-light duration in advance (Figure \ref{fig:set_phase_duration}). 
    This design enables the direct allocation of green time to different traffic flows, reducing the need for frequent switching decisions.
    \begin{figure}[H]
        \centering
        \includegraphics[width=0.95\linewidth]{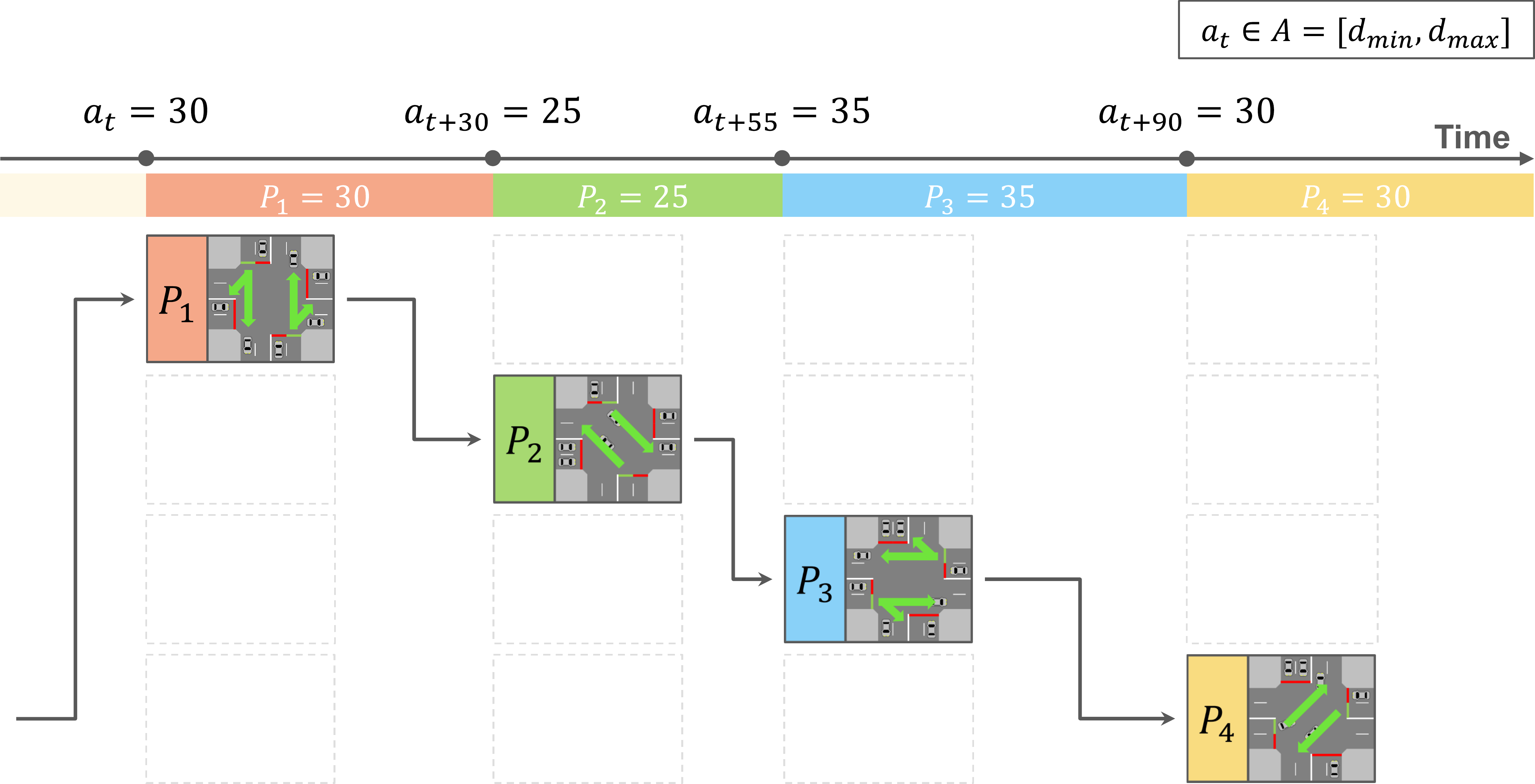}
        \caption{Illustration of the \emph{Set Phase Duration} action design.}
        \label{fig:set_phase_duration}
    \end{figure}
\end{itemize}
Together, these three types cover different aspects of signal control—phase selection, phase extension, and green-time allocation—providing diverse strategies for adaptive traffic management.

\begin{figure*}[t]
    \centering
    \includegraphics[width=0.9\linewidth]{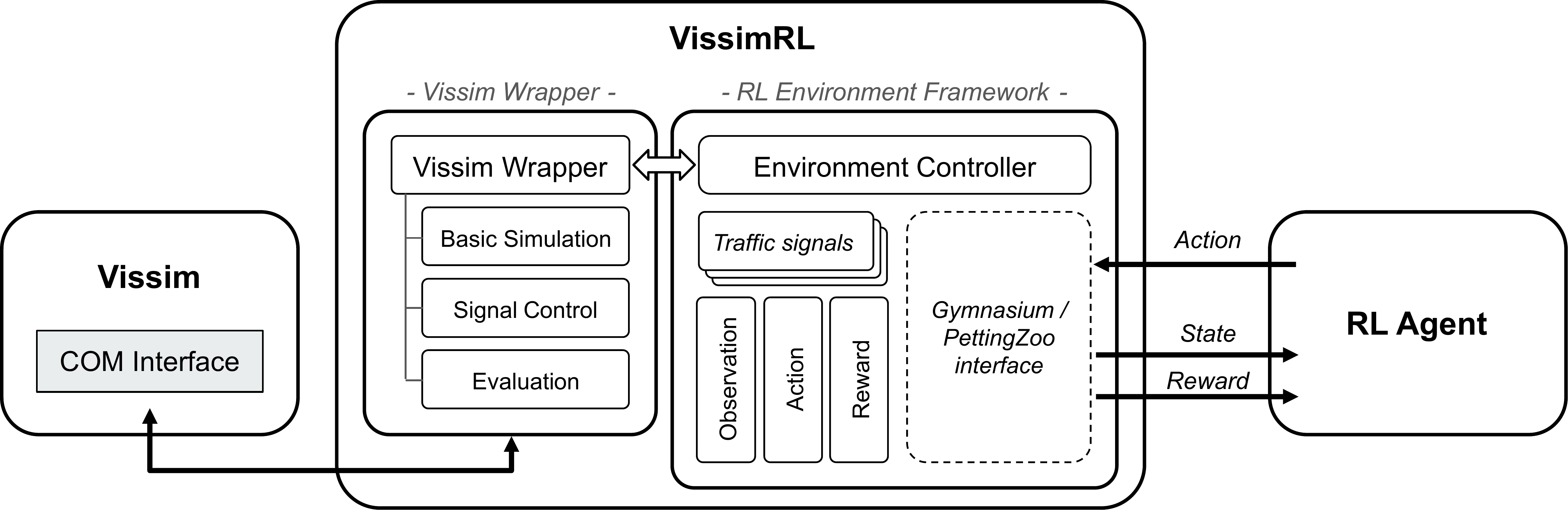}
    \caption{Architecture of the VissimRL framework.}
    \label{fig:vissimrl_architecture}
\end{figure*}

\section{System Design of VissimRL}
The proposed VissimRL framework consists of two core components, the Vissim Wrapper and the RL Environment Framework, as shown in Figure \ref{fig:vissimrl_architecture}. 
The Vissim Wrapper interacts with Vissim through the COM interface, covering essential functions such as simulation control, signal operation, and performance evaluation. 
The RL Environment Framework translates the interactions between Vissim and RL agents into RL components, namely states, actions, and rewards, while providing standardized interfaces (Gymnasium / PettingZoo APIs) to ensure seamless integration with mainstream RL algorithms. 

\subsection{Vissim Wrapper}
The Vissim Wrapper is built on top of Vissim COM interface to simplify simulator interactions and lower the development barrier for RL integration. 
While the COM interface provides direct access to internal objects and methods, its complexity is evident from the Vissim 2025 document \cite{fellendorfMicroscopicTrafficFlow2010}, which lists 276 object types and 4,164 attributes. 
The hierarchical structure and intricate attribute-access mechanism often result in high development costs. 
To address this, the Vissim Wrapper encapsulates low-level COM calls into an intuitive Python API and supports configuration-based setup of simulation environments, thereby reducing the complexity of implementation. 
It is also designed to be RL-friendly, enabling dynamic phase control and real-time access to traffic states and performance indicators.

The advantage of the Vissim Wrapper is not only in hiding the complexity of COM, but also in its reusability and maintainability. 
It can be applied across projects, and only minor adjustments are required when Vissim updates to ensure stable operation. 
The architecture of this component comprises three modules, whose functional designs and main interfaces are introduced in the following subsections.

\subsubsection{Basic Simulation}
The Basic Simulation module is responsible for creating and controlling the overall simulation environment, enabling rapid initialization through simple configuration. 
Users can specify the network and signal files, along with simulation parameters, which are automatically loaded and initialized by the Wrapper (see Appendix~\ref{appendix:wrapper_config} for detailed configuration parameters). 
During execution, this module provides intuitive time-control operations, ensuring both reproducibility and flexibility. 
The simulation can be advanced according to the defined simulation resolution (e.g., 10 steps per second), allowing precise control of the simulation process.
Its main functions are summarized in Table \ref{tab:basic_functions}.

\begin{table}[h]
    \centering
    \begin{threeparttable}
    \caption{Basic simulation functions and their descriptions.}
    \label{tab:basic_functions}
    \begin{tabularx}{\linewidth}{lX}
        \toprule
        \textbf{Function} & \textbf{Description} \\ \midrule
        \texttt{vissim.start()} & Start simulation \\ 
        \texttt{vissim.stop()} & Stop simulation \\ 
        \texttt{vissim.step()} & Advance the simulation by one resolution step \\ 
        \texttt{vissim.step\_one\_sec()} & Advance the simulation by one second \\ 
        \texttt{vissim.run\_until(t)} & Run the simulation till the specified time \texttt{t} \\ \bottomrule
    \end{tabularx}
    \begin{tablenotes}
        \footnotesize
        \item \textit{Note.} Use \texttt{vissim = VissimWrapper(**config)} to create a simulation environment.
    \end{tablenotes}
    \end{threeparttable}
\end{table}

\subsubsection{Signal control}
The Signal Control module manages intersection signal systems and synchronizes with agent decisions in real time. 
It supports querying current phase states and issuing phase-switching commands, ensuring consistency between simulation status and RL agent control. 
Through a unified and user-friendly interface, users can modify signal logic without delving into Vissim's low-level signal object structures. Key functions are listed in Table \ref{tab:signal_control_functions}.

\begin{table*}[t]
    \centering
    \begin{threeparttable}
    \caption{Signal control functions and their descriptions.}
    \label{tab:signal_control_functions}
    \begin{tabularx}{\linewidth}{p{6.5cm}X}
        \toprule
        \textbf{Function} & \textbf{Description} \\ \midrule
        \texttt{sc.set\_ts\_phase(ts\_id, phase\_state)} & Set the current \texttt{phase\_state}\tnote{1} of the specified intersection \texttt{ts\_id} \\ 
        \texttt{sc.get\_ts\_phase(ts\_id)} & Retrieve the current phase state of the specified intersection \texttt{ts\_id} in real time \\ \bottomrule
    \end{tabularx}
    \begin{tablenotes}
        \footnotesize
        \item \textit{Note.} Use \texttt{vissim.sc} for straightforward phase querying and control.
        \item[1] The term \texttt{phase\_state} refers to the signal status of all traffic movements within a phase (e.g., \texttt{"GRRR"}, where \texttt{G} denotes green and \texttt{R} denotes red).
    \end{tablenotes}
    \end{threeparttable}
\end{table*}

\subsubsection{Evaluation}
The Evaluation module provides real-time calculation of traffic performance indicators, serving as quantitative feedback for the RL environment. 
It supports both network-level and intersection-level metrics: network-level indicators include travel time, distance, and delay, while intersection-level indicators provide local information such as vehicle counts and queue lengths. 
The main functions are shown in Table \ref{tab:evaluation_functions}. 
These performance measures are further transformed into standardized RL-TSC indicators, which serve as states or rewards for the agent (e.g., the change in delay between decision steps).

\begin{table*}[h]
    \centering
    \begin{threeparttable}
    \caption{Evaluation functions and their descriptions.}
    \label{tab:evaluation_functions}
    \begin{tabularx}{\linewidth}{p{6.5cm}X}
        \toprule
        \textbf{Function} & \textbf{Description} \\ \midrule
        \texttt{eval.total\_travel\_time} & Total travel time of all vehicles in the network  \\ 
        \texttt{eval.total\_travel\_distance} & Total travel distance of all vehicles in the network \\ 
        \texttt{eval.total\_delay} & Total delay time for all vehicles in the network \\ 
        \texttt{eval.total\_iwaiting\_time} & Total waiting time for all vehicles in the network (internal) \\ 
        \texttt{eval.total\_bwaiting\_time} & Total waiting time for all vehicles at network entry points (boundary) \\ 
        \texttt{eval.total\_arrived} & Total number of vehicles that have arrived at their destinations and have left the network\\ 
        \texttt{eval.get\_num\_vehicles(ts\_id)} & Returns the current number of vehicles in each lane at the specified intersection (\texttt{ts\_id}) \\ 
        \texttt{eval.get\_queue\_length(ts\_id)} & Returns the current queue length in each lane at the specified intersection (\texttt{ts\_id}) \\ \bottomrule
    \end{tabularx}
    \begin{tablenotes}
        \footnotesize
        \item \textit{Note.} Use \texttt{vissim.eval} for accessing real-time traffic performance metrics, which can be further processed into RL-TSC indicators.
    \end{tablenotes}
    \end{threeparttable}
\end{table*}

\subsection{RL Environment Framework}
The proposed RL Environment Framework is specifically developed for TSC and is tightly integrated with the Vissim Wrapper. 
It supports the definition of states, actions, and rewards required for RL, and implements standardized interfaces following the Gymnasium and PettingZoo APIs to ensure compatibility with popular RL libraries such as Stable-Baselines3 and RLlib. 
The framework adopts a modular design, providing built-in implementations of commonly used state, action, and reward functions in RL-TSC, while offering base classes for easy inheritance and extension, allowing researchers to customize components according to experimental needs.

\begin{figure}[h]
    \centering
    \includegraphics[width=1\linewidth]{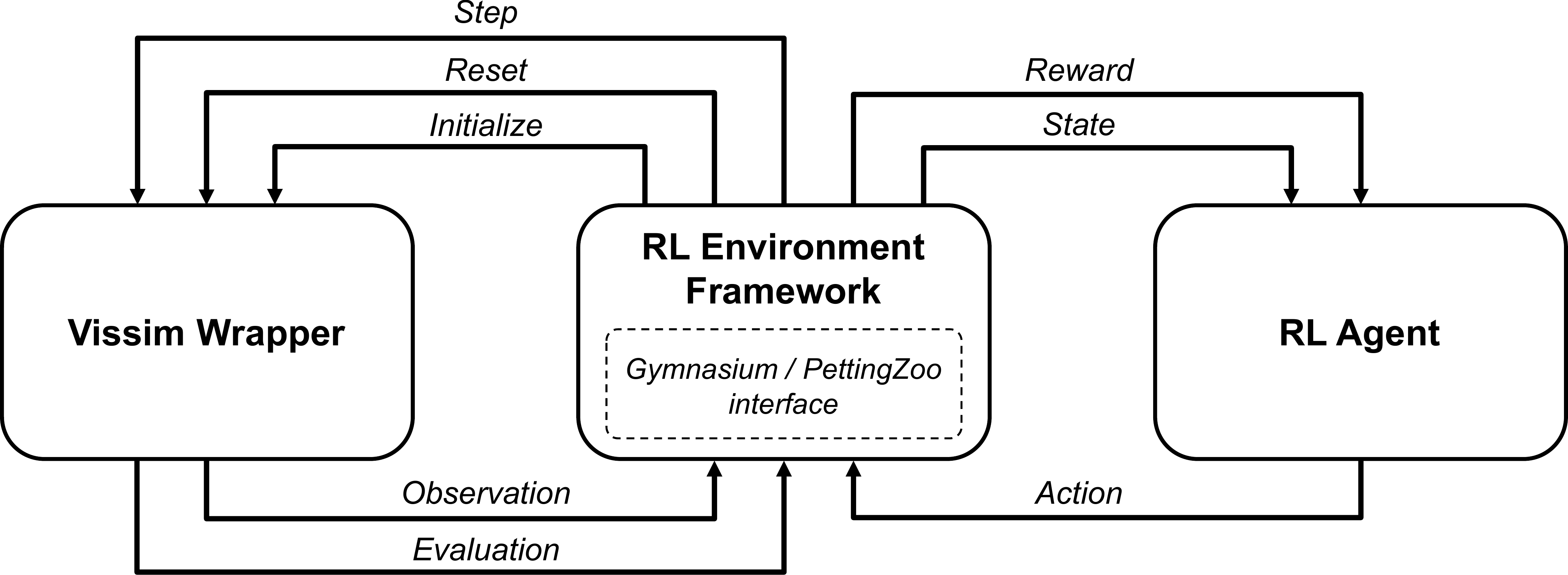}
    \caption{Operation flow of the VissimRL environment framework with standard API interfaces.}
    \label{fig:rl_env_framework}
\end{figure}

The operation flow of the RL Environment Framework, aligned with standard RL APIs, is illustrated in Figure \ref{fig:rl_env_framework}:
\begin{enumerate}
    \item \textit{Initialize}: Before training, the environment loads configuration files, builds the Vissim simulation, and initializes the control and evaluation modules.
    \item \textit{Reset}: At the beginning of each episode, the simulation is reset, and the initial observation is converted into a state vector usable by the agent.
    \item \textit{Step}: At each decision step, the agent selects an action based on the state vector. The framework maps the action into traffic signal control commands applied to Vissim, advances the simulation to the next decision point, computes the corresponding reward, and returns it together with the updated state.
\end{enumerate}
The step process repeats until the end of the simulation horizon, after which the environment can be reset for the next training episode, forming a complete RL interaction loop. This design ensures synchronization between decisions and simulation states, supports both single- and multi-agent training, and allows existing RL algorithms to be directly applied to high-fidelity traffic control experiments.

\subsubsection{Environment Controller}
In the RL Environment Framework, the Environment Controller serves as the central management layer, responsible for scenario loading, parameter initialization, and simulation time progression. 
It connects the Vissim simulator with agent policies through standardized RL interfaces (Figure \ref{fig:rl_env_framework}). 
At the beginning of each training episode, the environment executes the \texttt{\small reset()} function to reinitialize the simulator state and signal phases, followed by a warm-up procedure to ensure a stable traffic distribution. 
During each decision step, the RL agent invokes the \texttt{\small step(actions)} function, which sends the \texttt{\small actions} to the corresponding Traffic Signals modules within VissimRL, advancing the simulation to the next decision point, and then returns updated observations, rewards computed from performance measures, as well as termination, truncation, and auxiliary information. 
The core functionalities and interface definitions are summarized in Table \ref{tab:env_controller_functions}.

In addition, the module provides a simple and standardized interface, allowing users to instantiate the RL environment with a single line of code:
\begin{itemize}
    \item For single-agent scenarios: {\small\texttt{gym.make(**config)}} 
    \item For multi-agent scenarios: {\small\texttt{parallel\_env(**config)}}
\end{itemize}
This design significantly lowers the usage barrier and establishes a highly reproducible experimental foundation for subsequent RL training (see Appendix~\ref{appendix:rlenv_config} for the complete list of configuration parameters).

\begin{table}[h]
    \centering
    \begin{threeparttable}
    \caption{Environment controller functions and their descriptions.}
    \label{tab:env_controller_functions}
    \begin{tabularx}{\linewidth}{lX}
        \toprule
        \textbf{Function} & \textbf{Description} \\ \midrule
        \texttt{reset()} & Resets the network, initializes all traffic signals, runs the warmup procedure, and returns the initial observation and info \\ 
        \texttt{step(actions)} & Sends \texttt{actions} to the corresponding traffic signal controllers, advances the simulation to the next decision point, and returns the next observations, rewards, terminations, truncations, and infos \\ 
        \bottomrule
    \end{tabularx}
    \end{threeparttable}
\end{table}

\subsubsection{Traffic signals}
The Traffic Signals module assigns each intersection an independent controller to manage phase sequences and performance indicators, ensuring that in multi-intersection scenarios, decisions can be made locally with corresponding observation, action, and reward functions. 
This architecture supports flexible multi-agent designs where agents act based on their own traffic conditions. 
To ensure safety and stability, the system automatically inserts yellow and all-red intervals during phase transitions and enforces minimum and maximum green durations to avoid excessive switching. 
These mechanisms provide a reliable environment for RL-based control strategies. 
The function to support the signal control is  \texttt{\small ts.set\_next\_phase(phase\_id, duration)} as in Table \ref{tab:traffic_signal_functions}, where \texttt{\small ts} is an given intersection.

\begin{table}[h]
    \centering
    \begin{threeparttable}
    \caption{Traffic signal functions and their descriptions.}
    \label{tab:traffic_signal_functions}
    \begin{tabularx}{\linewidth}{lX}
        \toprule
        \textbf{Function} & \textbf{Description} \\ \midrule
        \texttt{\makecell[lt]{ts.set\_next\_phase(\\\hspace{2em}phase\_id, duration)}} & Set the next phase \texttt{phase\_id} and its \texttt{duration} for the upcoming step. \\ 
        \bottomrule
    \end{tabularx}
    \end{threeparttable}
\end{table}

\subsubsection{Three RL elements}
The RL Elements consist of three pluggable modules: Observation, Action, and Reward, which form the core logic of the RL environment. 
In each module, a callable class is implemented to support inheritance and extension, allowing researchers to flexibly define functionalities for different traffic scenarios and experimental requirements.
    
\textit{The Observation module} generates state vectors for RL agents and supports both local and global observations. 
The default local observation features include traffic characteristics such as phase state, queue length, and traffic density, while global observation aggregates multi-intersection information into a single input for coordinated multi-agent control. 
The detailed observation functions are listed in Table \ref{tab:observation_functions}.

\begin{table}[h]
    \centering
    \begin{threeparttable}
    \caption{Observation functions and their descriptions.}
    \label{tab:observation_functions}
    \begin{tabularx}{\linewidth}{lX}
        \toprule
        \textbf{Callable Class} & \textbf{Description} \\ \midrule
        \texttt{LocalObservationFunction}\tnote{1} & Generates observation vectors for a single intersection \\ 
        \quad - Phase state & Current active signal phase \\ 
        \quad - Min green & Whether the current phase is within minimum green constraint \\ 
        \quad - Density & Ratio of vehicles on a lane to its maximum capacity \\ 
        \quad - Queue length & Number of vehicles waiting in queue \\ 
        \quad - Elapsed time & Time elapsed since the last signal change \\ 
        \texttt{GlobalObservationFunction} & Aggregates local observations from all controlled intersections \\ 
        \bottomrule
    \end{tabularx}
    \begin{tablenotes}
        \footnotesize
        \item \textit{Note.} Observation classes implement \texttt{\_\_call\_\_()} and \texttt{observation\_space()} for compatibility with RL frameworks.
        \item[1] The listed items are the internal features included in its observation vector.
    \end{tablenotes}
    \end{threeparttable}
\end{table}

\textit{The Action module} defines the signal control strategies available to RL agents. 
It currently provides three common RL-TSC actions in classes of \texttt{\small ChooseNextPhase}, \texttt{\small SwitchNextOrNot}, \texttt{\small SetPhaseDuration}, corresponding to those three actions described in Section~\ref{subsubsec:actions}.
These actions are mapped through the Traffic Signals module to Vissim's control logic, ensuring that agent decisions are immediately reflected in the simulation environment. 
The functions and parameter settings are summarized in Table \ref{tab:action_functions}.

\begin{table}[h]
    \centering
    \begin{threeparttable}
    \caption{Action functions and their descriptions.}
    \label{tab:action_functions}
    \begin{tabularx}{\linewidth}{lX}
        \toprule
        \textbf{Callable Class} & \textbf{Description} \\ \midrule
        \texttt{ChooseNextPhase} & \makecell[tl]{Choose any phase as the next phase;\\ args: \texttt{delta\_time} (default: 5s)} \\ 
        \texttt{SwitchNextOrNot} & \makecell[tl]{Decide whether to switch to the next phase;\\ args: \texttt{delta\_time} (default: 5s)} \\ 
        \texttt{SetPhaseDuration} & Set the duration of the new phase before activation \\ 
        \bottomrule
    \end{tabularx}
    \begin{tablenotes}
        \footnotesize
        \item \textit{Note.} Action classes implement \texttt{\_\_call\_\_()} and \texttt{action\_space()} for compatibility with RL frameworks.
    \end{tablenotes}
    \end{threeparttable}
\end{table}

\textit{The Reward module} computes traffic performance during each decision step and supports multi-objective weighted combinations to balance indicators such as internal waiting time, boundary waiting time, travel time. 
The details of the reward functions are presented in Table \ref{tab:reward_functions}.

\begin{table}[h]
    \centering
    \begin{threeparttable}
    \caption{Reward functions and their descriptions.}
    \label{tab:reward_functions}
    \begin{tabularx}{\linewidth}{lX}
        \toprule
        \textbf{Callable Class} & \textbf{Description} \\ \midrule
        \texttt{DefaultRewardFunction} & Computes a weighted combination of multiple traffic performance metrics \\ 
        \quad - Internal waiting time & Measures the time vehicles wait within the network \\ 
        \quad - Boundary waiting time & Measures the waiting time at network entry points \\ 
        \quad - Travel time & Computes the total travel time of vehicles \\ 
        \quad - Throughput & Counts the number of vehicles passing through intersections \\ 
        \quad - Speed & Calculates the average vehicle speed \\ 
        \quad - Delay & Quantifies total delay compared to free-flow conditions \\ 
        \bottomrule
    \end{tabularx}
    \begin{tablenotes}
        \footnotesize
        \item \textit{Note.} Reward classes implement \texttt{\_\_call\_\_()} for real-time reward computation and support configurable weights for multi-objective designs.
    \end{tablenotes}
    \end{threeparttable}
\end{table}

With the help of a configuration file, all modules are automatically instantiated for each Traffic Signal upon environment initialization, ensuring high flexibility and scalability in both single-agent and multi-agent training scenarios. 

\section{Experiments}

The experimental design of this study aims to comprehensively evaluate the effectiveness and applicability of the proposed VissimRL framework, divided into two complementary levels according to the system architecture.
First, in the assessment of the Vissim Wrapper, the focus is on quantifying its impact on development effort and runtime efficiency. 
By comparing the source lines of code (SLOC) required for equivalent control tasks using Raw COM versus the Vissim Wrapper, the study evaluates the practical benefits of the high-level API in reducing development cost. 
In addition, per-step latency and simulation-step throughput are measured under different standardized workloads to examine the performance implications of the abstraction layer introduced for improved usability.

Second, in the validation of the RL Environment Framework, the experiments are designed to examine the effectiveness and applicability of the proposed VissimRL framework under different TSC scenarios. 
The framework integrates Vissim with standardized RL interfaces to evaluate performance across three action designs ({\small\texttt{ChooseNextPhase}}, {\small\texttt{SwitchNextOrNot}}, and {\small\texttt{SetPhaseDuration}}) in both single- and multi-agent settings. 
To further assess real-world applicability, a benchmark case with fixed-time control is selected, enabling quantitative evaluation of RL-based improvements within the VissimRL framework under realistic traffic conditions.

\subsection{Experimental Scenarios}
The experimental scenarios are divided into synthetic and real-world cases. 
The synthetic scenarios use artificially constructed networks and synthetic traffic demand, including a single-intersection network (for single-agent control) and a three-intersection arterial network (for multi-agent control). 
The real-world scenario is based on a network of five signalized intersections near the Dayuan Interchange in Taoyuan, Taiwan, using one hour of peak-period traffic data to validate the proposed method under realistic conditions.

\subsubsection{Synthetic scenario}
\begin{figure}[h]
    \centering
    \begin{subfigure}[b]{0.24\linewidth}
        \centering
        \includegraphics[height=2.2cm]{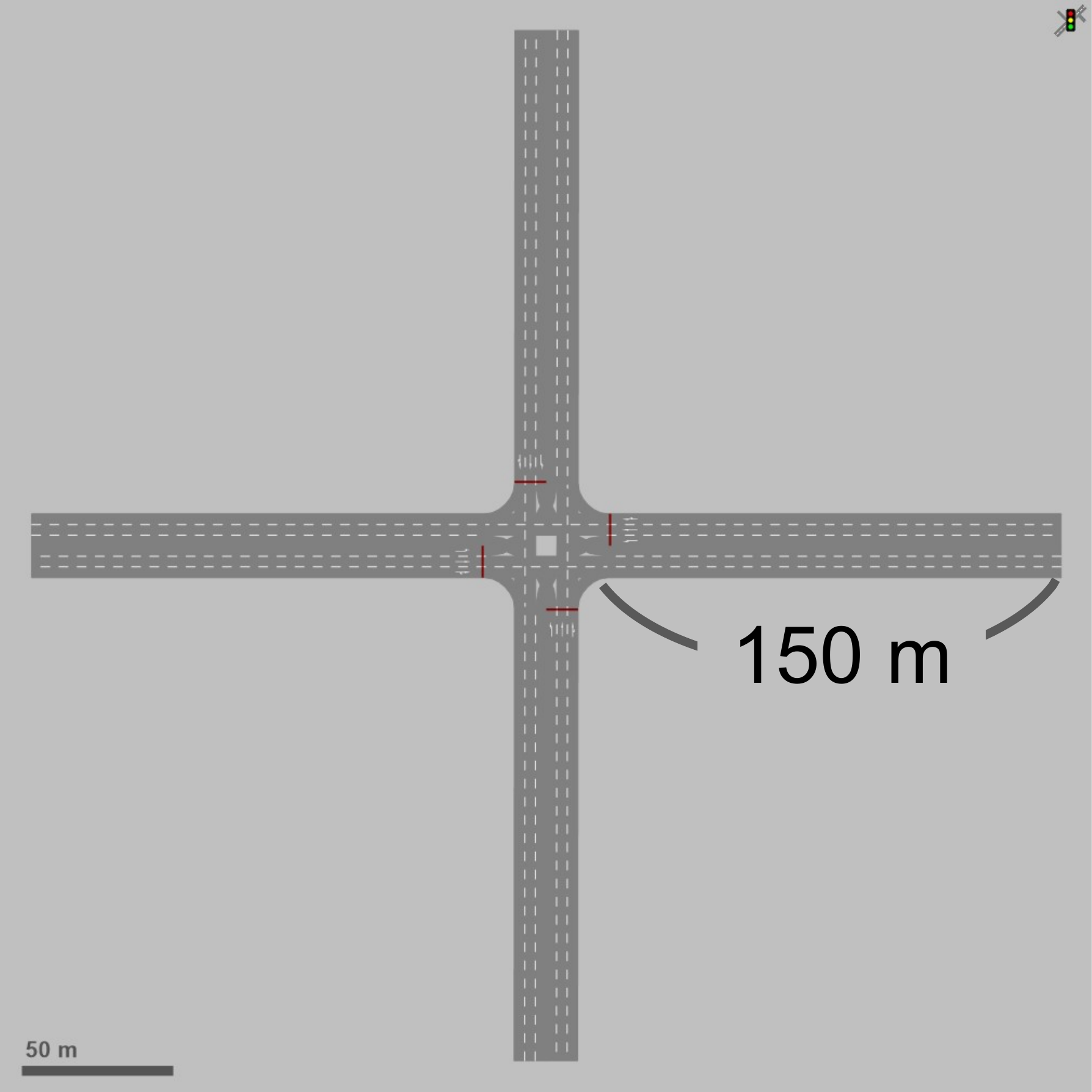}
        \caption{}
        \label{fig:single_intersection}
    \end{subfigure}
    \hfill
    \begin{subfigure}[b]{0.74\linewidth}
        \centering
        \includegraphics[height=2.2cm]{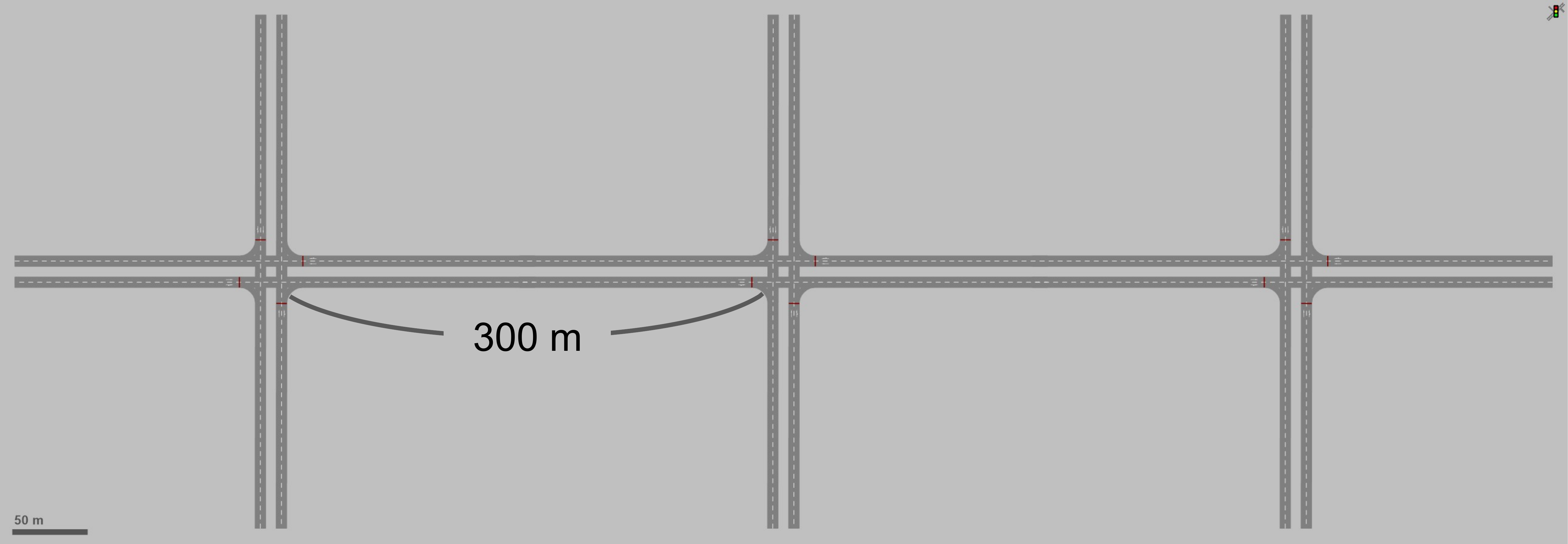}
        \caption{}
        \label{fig:arterial_three}
    \end{subfigure}
    \caption{Synthetic traffic networks: (a) Single-Intersection and (b) Arterial-3, an arterial network with three intersections.}
    \label{fig:synthetic_networks}
\end{figure}

In the synthetic experiments, two network structures are designed to correspond to single-agent and multi-agent RL control. 
The first is the Single-Intersection network as in Figure \ref{fig:single_intersection}, where each boundary link is 150 meters long, representing a symmetric and relatively simple traffic environment. 
The second is the Arterial-3 network as in Figure \ref{fig:arterial_three}, consisting of three consecutive intersections connected by 300-meter links, used to evaluate coordination in multi-intersection scenarios.

The traffic demand and signal settings are summarized in Table \ref{tab:synthetic_traffic_config}. 
In the Single-Intersection network, directional imbalance is introduced by assigning higher east–west flows compared to north–south flows, with unequal turning ratios to reflect realistic directional differences. 
The signal plan includes four phases: east–west through (EW), east–west left (EW\_L), north–south through (NS), and north–south left (NS\_L). 
For the Arterial-3 network, the design focuses on progression and coordination along the east–west arterial, which also carries higher flows than the side streets. 
The intersections are restricted to through movements only, with a simplified two-phase signal plan designed to highlight arterial coordination effects. 
The vehicle composition includes only cars, with a uniform desired speed of 50 km/h to ensure comparability and stable evaluation across networks.

\begin{table}[h]
    \centering
    \begin{threeparttable}
    \caption{Configurations for traffic data.}
    \label{tab:synthetic_traffic_config}
    \begin{tabular}{llll}
        \toprule
        \makecell[l]{\textbf{Network}} & 
        \makecell[l]{\textbf{Arrival rate}\\\textbf{(vehicles/h/road)}} & 
        \makecell[l]{\textbf{Turn ratio}\\\textbf{(L : S : R)}} & 
        \makecell[l]{\textbf{Phases}} \\ \midrule
        Single-Intersection & \makecell[l]{EW-flow : 1200\\NS-flow : 800} & \makecell[l]{0.1 : 0.8 : 0.1} & \makecell[l]{EW, EW\_L,\\ NS, NS\_L} \\[2.3ex]
        Arterial-3 & \makecell[l]{Arterial : 1200\\Side-street : 800} & \makecell[l]{0 : 1 : 0} & \makecell[l]{EW, NS} \\ \bottomrule
    \end{tabular}
    \begin{tablenotes}
        \footnotesize
        \item \textit{Note.} All networks use cars only, with a desired speed of 50 km/h.  
        \item NS = North-South; EW = East-West; L = Left-turn phase.  
        \item L:S:R = Left : Straight : Right turn ratio.
    \end{tablenotes}
    \end{threeparttable}
\end{table}

\subsubsection{Real-world scenario}
\begin{figure}[h]
    \centering
    \begin{subfigure}{1\linewidth}
        \centering
        \includegraphics[width=\linewidth]{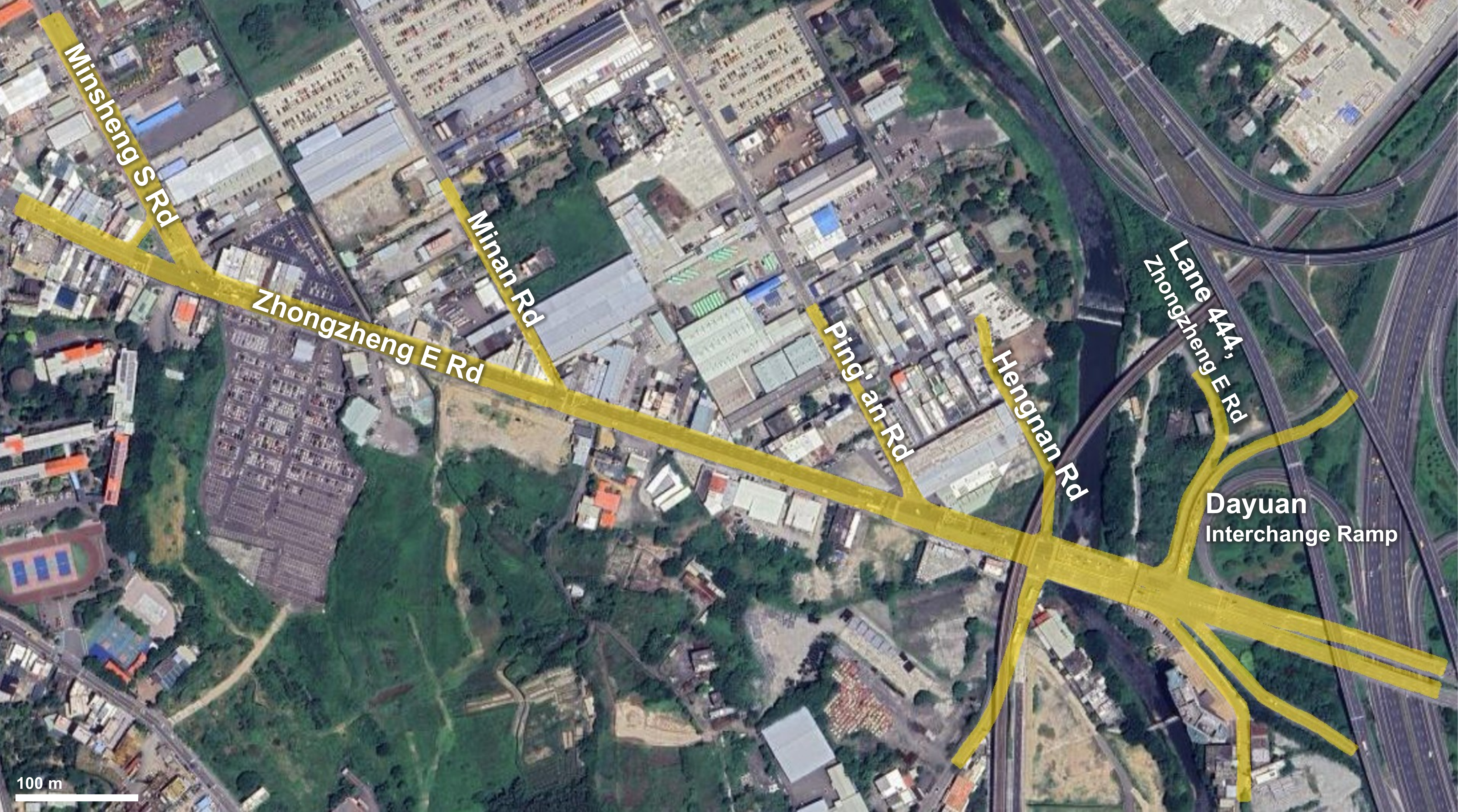}
        \caption{Aerial photograph of the Dayuan Interchange.}
        \label{fig:dayuan_aerial}
    \end{subfigure}
    \vskip\baselineskip
    \begin{subfigure}{1\linewidth}
        \centering
        \includegraphics[width=\linewidth]{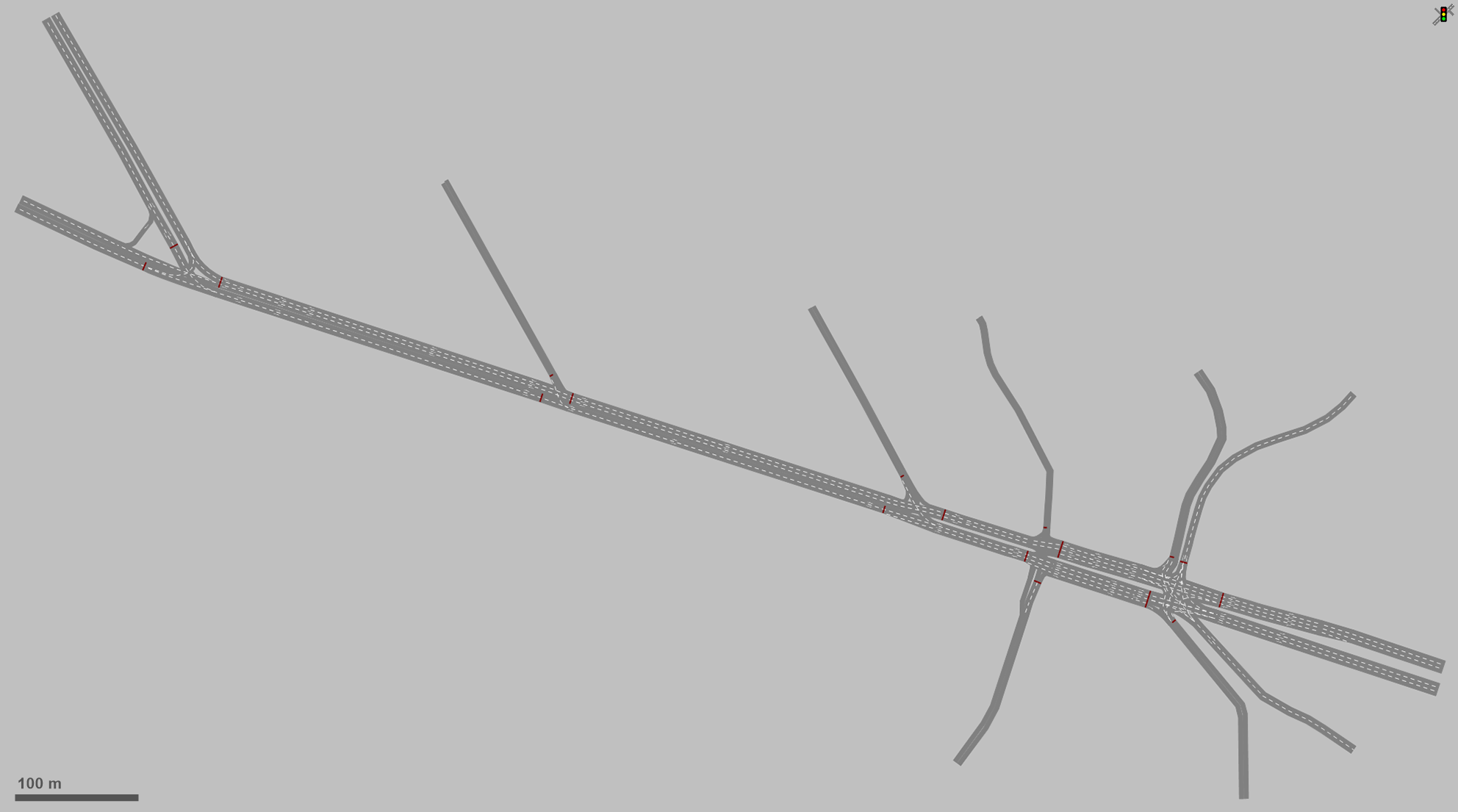}
        \caption{Corresponding simulation network in Vissim.}
        \label{fig:dayuan_simulation}
    \end{subfigure}
    \caption{Real-world scenario of the Dayuan Interchange: (a) aerial photograph and (b) corresponding simulation network.}
    \label{fig:dayuan_scenario}
\end{figure}

For the real-world experiments, a network of five signalized intersections near the Dayuan Interchange in Taoyuan, Taiwan, is selected to evaluate the applicability and effectiveness of the VissimRL framework under realistic traffic conditions. 
This area includes a major arterial and several side streets, featuring significant flow imbalances and complex intersection structures. 
The real-world network and its simulated counterpart are shown in Figure \ref{fig:dayuan_scenario}, where (a) depicts an aerial view of the study area and (b) illustrates the corresponding simulation network.

Table \ref{tab:real_traffic_config} details the inflows and signal phase settings of the five intersections. 
The simulation uses one hour of observed peak-period traffic demand, with a vehicle composition of cars, buses, and motorcycles, and a desired speed of 50 km/h. 
Overall, traffic volumes on the arterial are significantly higher than those on the side streets, reflecting the imbalanced conditions commonly observed in practice. 
This configuration captures the complexity of real-world traffic conditions and serves to examine VissimRL's adaptability and optimization capability under varying demand pressures and signal complexities.

\begin{table}[h]
    \centering
    \begin{threeparttable}
    \caption{Configurations for real-world traffic data.}
    \label{tab:real_traffic_config}
    \begin{tabular}{lll}
        \toprule
        \makecell[l]{\textbf{Network}} &
        \makecell[l]{\textbf{Arrival rate}\\\textbf{(vehicles/h)}} &
        \makecell[l]{\textbf{Phases}\\\textbf{(number)}} \\
        \midrule
        \makecell[l]{Dayuan\\(5-Intersections)} &
        \makecell[l]{Arterial : 3573\\Side-street : 2420} &
        - \\
        \addlinespace[3pt] 
        \hdashline[3pt/1.5pt]
        \addlinespace[3pt] 
        \ \textit{- Intersection-1} &
        \makecell[lt]{Minsheng S Rd : 1623\\Zhongzheng E Rd : 869} &
        2 \\
        \ \textit{- Intersection-2} &
        \makecell[lt]{Minan Rd : 46} &
        2 \\
        \ \textit{- Intersection-3} &
        \makecell[lt]{Ping' an Rd : 498} &
        2 \\
        \ \textit{- Intersection-4} &
        \makecell[lt]{Hengnan Rd : 396} &
        3 \\
        \ \textit{- Intersection-5} &
        \makecell[lt]{Zhongzheng E Rd : 1950\\Dayuan Interchange Ramp : 412\\
        Lane 444, Zhongzheng E Rd : 199} &
        4 \\
        \bottomrule
    \end{tabular}
    \begin{tablenotes}
        \footnotesize
        \item \textit{Note.} Vehicle composition includes cars, buses, and scooters (with a desired speed of 50 km/h).
    \end{tablenotes}
    \end{threeparttable}
\end{table}

\subsection{Wrapper Evaluation}
This section evaluates the performance of the Vissim Wrapper in terms of development effort and runtime efficiency. 
To ensure consistency and comparability, we define three standardized benchmark workloads. 
They represent progressively more complex levels of interaction, ranging from basic simulation advancement to the combination of signal control and performance evaluation.

\begin{itemize}
    \item W0 — Step-only\\
    Advances the simulation by time steps only. This workload reflects the theoretical lower bound of simulation execution and is primarily used to measure the baseline overhead of advancing simulation steps.
    \item W1 — Step + TS\\
    Extends W0 by adding traffic signal read/write operations. Every five seconds, the current phase is retrieved and switched to the next, representing a simplified signal control loop.
    \item W2 — Step + TS + Eval\\
    Builds on W1 by incorporating traffic state and performance queries. Queue length is used as the observation feature, and delay is adopted as the performance indicator, representing a simplified RL training loop.
\end{itemize}

Among these, W0 serves as a performance baseline, while W1 and W2 represent increasingly realistic control and learning interactions. 
The stepwise design of the workloads allows the impact of different task types on development effort and runtime efficiency to be distinguished, offering a structured way to assess the Vissim Wrapper's performance.

\subsubsection{Runtime Micro-benchmark}
To assess the runtime efficiency of the Vissim Wrapper, its performance is compared with Raw COM across standardized workloads.
The experiment was conducted on a Single-Intersection network with a simulation length of 3600 seconds and a resolution of 10 (36,000 simulation steps). 
Logging, visualization, and non-essential outputs were disabled to ensure that the measurements captured only the overhead of program calls. 
The evaluation considered two indicators: per-step latency, which measures the average execution time of each simulation step to reflect fine-grained computational overhead, and simulation-step throughput, representing the overall simulation efficiency.

The results are summarized in Tables \ref{tab:per_step_latency} and \ref{tab:step_throughput}.
For the W0 workload, which involves only simulation stepping, the Wrapper showed only a 4\% increase in per-step latency compared to Raw COM, indicating negligible overhead, while the simulation-step throughput was maintained at a comparable level. 

\begin{table}[h]
    \centering
    \begin{threeparttable}
    \caption{Per-step latency comparison between Raw COM and Wrapper for benchmark workloads.}
    \label{tab:per_step_latency}
    \begin{tabular}{lccc}
        \toprule
        \textbf{Workload} & 
        \makecell[c]{\textbf{Raw COM} \\\textbf{(ms/step)}} &  
        \makecell[c]{\textbf{Wrapper} \\\textbf{(ms/step)}} & 
        \textbf{Overhead (\%)} \\
        \midrule
        W0 Step-only & 0.787 {\scriptsize $\pm$ 0.005} & 0.819 {\scriptsize $\pm$ 0.006} & $+\ 4.041$ \% \\
        W1 Step+TS & 0.889 {\scriptsize $\pm$ 0.004} & 0.828 {\scriptsize $\pm$ 0.005} & $-\ 6.842$ \% \\
        W2 Step+TS+Eval & 0.967 {\scriptsize $\pm$ 0.004} & 0.848 {\scriptsize $\pm$ 0.005} & $-\ 12.267$ \% \\
        \bottomrule
    \end{tabular}
    \begin{tablenotes}
        \footnotesize
        \item \textit{Note.} Values represent mean $\pm$ standard deviation across five runs. Lower latency indicates better performance.
    \end{tablenotes}
    \end{threeparttable}
\end{table}

\begin{table}[h]
    \centering
    \begin{threeparttable}
    \caption{Simulation-step throughput comparison between Raw COM and Wrapper for benchmark workloads.}
    \label{tab:step_throughput}
    \begin{tabular}{>{\arraybackslash}m{2.5cm}
                    >{\centering\arraybackslash}m{1.5cm}
                    >{\centering\arraybackslash}m{1.5cm}
                    >{\centering\arraybackslash}m{1.3cm}}
        \toprule
        \textbf{Workload} & 
        \makecell[c]{\textbf{Raw COM} \\\textbf{(steps/s)}} &  
        \makecell[c]{\textbf{Wrapper} \\\textbf{(steps/s)}} & 
        \textbf{Ratio} \\
        \midrule
        W0 Step-only & 1270.496 & 1220.949 & 0.961 \\
        W1 Step+TS & 1125.551 & 1208.045 & 1.073 \\
        W2 Step+TS+Eval & 1034.438 & 1179.033 & 1.140 \\
        \bottomrule
    \end{tabular}
    \begin{tablenotes}
        \footnotesize
        \item \textit{Note.} Values represent the mean over five runs. Higher throughput indicates better performance.
    \end{tablenotes}
    \end{threeparttable}
\end{table}

Interestingly, in the more interactive W1 and W2 workloads, the Wrapper not only avoids performance degradation but also achieves lower per-step latency, reduced by 6.8\% and 12.3\%, respectively, along with higher throughput (1.07x and 1.14x).
The major reason is that the Wrapper substantially reduces the number of calls by batching operations and caching repeated queries. 
This reduction in call frequency accounts for the improved runtime performance despite the additional abstraction layer. 
This outcome is also evident from the analysis of COM API calls shown in Table \ref{tab:api_calls}. 
Overall, the Wrapper preserves a concise development interface while integrating performance-oriented mechanisms such as batching and caching, offering both usability and efficiency, and providing a reliable foundation for building RL environments.

\begin{table}[h]
    \centering
    \begin{threeparttable}
    \caption{API calls comparison between Raw COM and Wrapper for benchmark workloads.}
    \label{tab:api_calls}
    \begin{tabular}{lccc}
        \toprule
        \textbf{Workload} & 
        \makecell[c]{\textbf{Raw COM} \\\textbf{(API calls)}} &  
        \makecell[c]{\textbf{Wrapper} \\\textbf{(API calls)}} & 
        \textbf{Reduction (\%)} \\
        \midrule
        W0 Step-only & \textbf{36000} & \textbf{36000} & 0 \% \\
        W1 Step+TS & 41760 & \textbf{37442} & 10.340 \% \\
        W2 Step+TS+Eval & 51132 & \textbf{40322} & 21.141 \% \\
        \bottomrule
    \end{tabular}
    \end{threeparttable}
\end{table}

\subsubsection{Development Effort Benchmark}
To evaluate the benefits of the Vissim Wrapper for development, this paper adopts source lines of code (SLOC) as a quantitative metric. 
SLOC counts only effective code containing logic, excluding comments and blank lines, thereby reflecting the development effort required to complete equivalent tasks.

The benchmark task follows the W2 workload, which involves simulation stepping, signal control, and performance queries, representing the typical operations in an RL training process. 
The same task was implemented in two ways:
\begin{itemize}
  \item \textbf{Raw COM implementation:} \\{\small\texttt{win32com.client.Dispatch("Vissim.Vissim")}}
  \item \textbf{Wrapper implementation:} \\{\small\texttt{VissimWrapper(**config)}}
\end{itemize}
The first approach required manual handling of initialization, control, and performance evaluation, whereas the wrapper version achieved the same functionality through a concise high-level API provided by VissimRL.

Table \ref{tab:sloc_comparison} summarizes the SLOC comparison between the two approaches based on the task for W2. 
From the table, using the Raw COM for the implementation of the task required 207 lines of code, while using the Wrapper for the same task required only 32, representing an 84.5\% reduction. 
The W2 workload reflects a simplified example, yet even in this reduced setting, the Raw COM implementation involves complex object hierarchies and property configurations that impose considerable effort for development and maintenance. 
More importantly, the Wrapper also hides the details of caching and batching from users, as mentioned in the previous subsection. 
By abstracting these low-level operations, the Wrapper achieves more than an 80\% reduction in code for the task, effectively minimizing redundant programming effort and enabling researchers to reproduce or adjust experiments through simple configuration changes. 
This substantially improves development efficiency and iteration speed.

\begin{table}[h]
    \centering
    \begin{threeparttable}
    \caption{Comparison of SLOC between Raw COM and Wrapper implementations.}
    \label{tab:sloc_comparison}
    \begin{tabular}{p{2cm} p{2cm} p{2cm}}
        \toprule
        \textbf{Metric} & \textbf{Raw COM} & \textbf{Wrapper} \\
        \midrule
        SLOC & 207 & 32 \\
        \bottomrule
    \end{tabular}
    \end{threeparttable}
\end{table}

\subsection{RL Environment Evaluation}
All experiments in this section are conducted using the simulation parameters listed in Table \ref{tab:simulation_settings}, with a total duration of 4200 seconds including a 600-second warm-up period to stabilize traffic flow. 
Signal timing settings, such as green duration limits and safety intervals (yellow and all-red), follow realistic traffic control requirements to ensure feasibility and safety.

\begin{table}[h]
    \centering
    \begin{threeparttable}
    \caption{Experimental configurations for simulation.}
    \label{tab:simulation_settings}
    \begin{tabular}{p{5.5cm} p{2cm}}
        \toprule
        \textbf{Parameter} & \textbf{Value (s)} \\ \midrule
        Warmup & 600 \\
        Simulation time (incl. warmup) & 4200 \\
        Min green duration & 5 \\
        Max green duration & 120 \\
        Yellow change interval & 3 \\
        All-red interval & 2 \\ \bottomrule
    \end{tabular}
    \end{threeparttable}
\end{table}

The RL algorithm is implemented using Proximal Policy Optimization (PPO) provided by RLlib, chosen for its advantages in continuous control and stable policy updates. 
For the observation design, the single-agent setting utilizes the {\small\texttt{LocalObservationFunction}}, which considers localized traffic states at the target intersection (as defined in Table \ref{tab:observation_functions}). In contrast, the multi-agent setting employs the {\small\texttt{GlobalObservationFunction}}, which integrates local observation vectors from multiple intersections into a global traffic state to support coordinated decision-making.

The reward function is based on the {\small\texttt{DefaultRewardFunction}} (see Table \ref{tab:reward_functions}) with a multi-objective weighted design to balance different traffic performance indicators. 
Internal waiting time and boundary waiting time penalize congestion inside intersections and at network boundaries, respectively. 
Throughput encourages higher vehicle volumes within a unit of time, while speed promotes smooth traffic flow by discouraging low-speed conditions. 
This weighting scheme provides an example of balancing delay reduction, efficiency improvement, and traffic smoothness. 
The detailed parameter settings are summarized in Appendix~\ref{appendix:rl_settings}.


For performance evaluation, three commonly used RL-TSC metrics \cite{noaeenReinforcementLearningUrban2022} are selected: delay, travel time, and waiting time. 
These indicators are used to examine the efficiency of different control strategies and to validate the applicability of the proposed system across various scenarios.


\subsubsection{Performance Validation}
In this experiment, training was conducted under two scenarios: Single-Intersection and Arterial-3, using three action designs implemented in our framework: {\small\texttt{ChooseNextPhase}}, {\small\texttt{SwitchNextOrNot}}, and {\small\texttt{SetPhaseDuration}}. 
For all evaluation metrics, lower values indicate better control performance. 
To observe performance dynamics during learning, the variations of each metric over training steps were plotted, providing an intuitive view of the convergence trends across different action designs and scenarios.

\begin{figure}[h]
    \centering
    \begin{subfigure}[t]{1\linewidth}
        \centering
        \includegraphics[width=\linewidth]{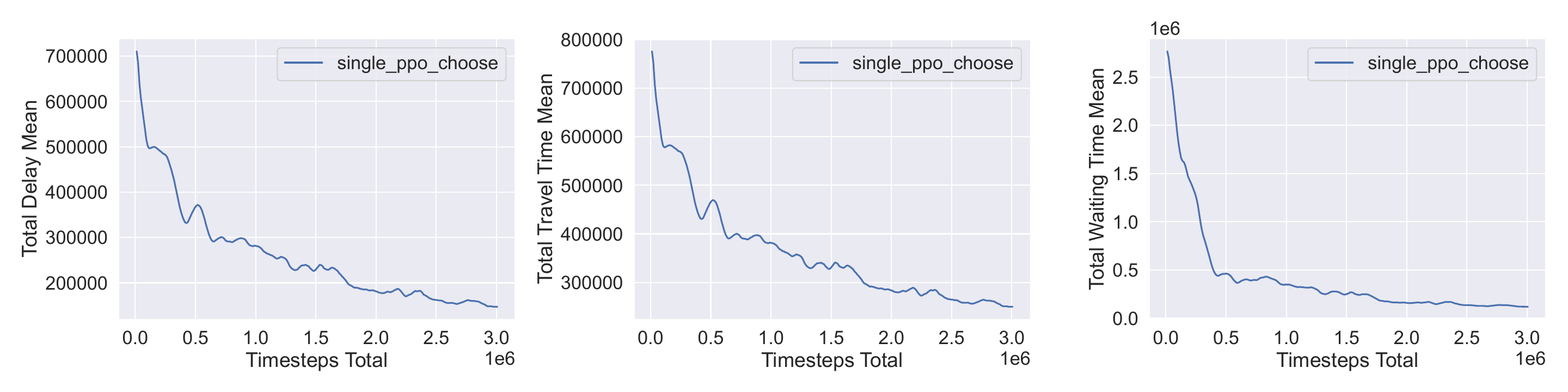}
        \caption{Single intersection}
        \label{fig:single_choose}
    \end{subfigure}

    \vspace{6pt}

    \begin{subfigure}[t]{1\linewidth}
        \centering
        \includegraphics[width=\linewidth]{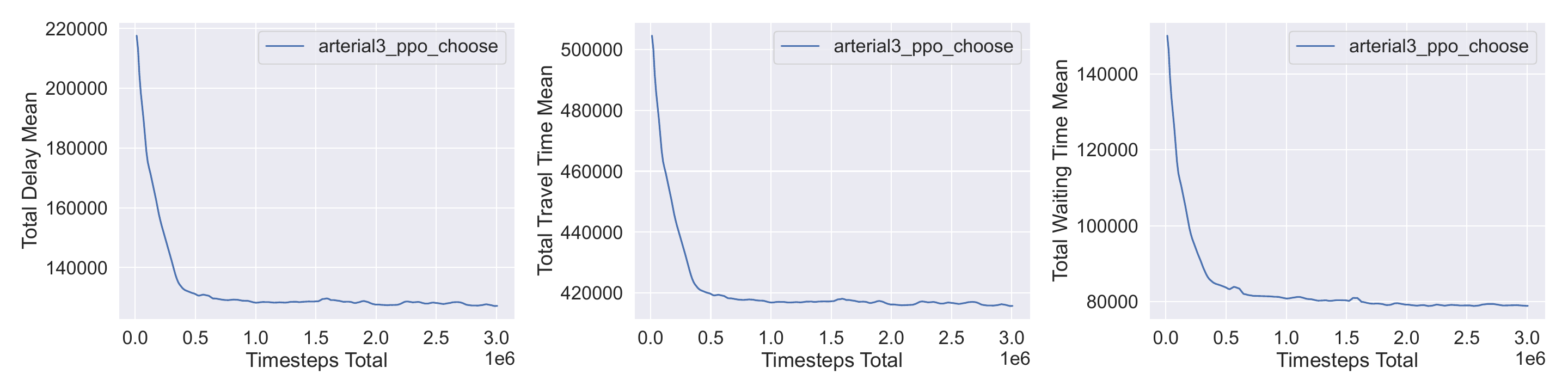}
        \caption{Arterial-3}
        \label{fig:arterial_choose}
    \end{subfigure}

    \caption{Training performance of {\small\texttt{ChooseNextPhase}} in different scenarios (top: Single; bottom: Arterial-3).}
    \label{fig:results_choose}
\end{figure}

\begin{figure}[h]
    \centering
    \begin{subfigure}[t]{1\linewidth}
        \centering
        \includegraphics[width=\linewidth]{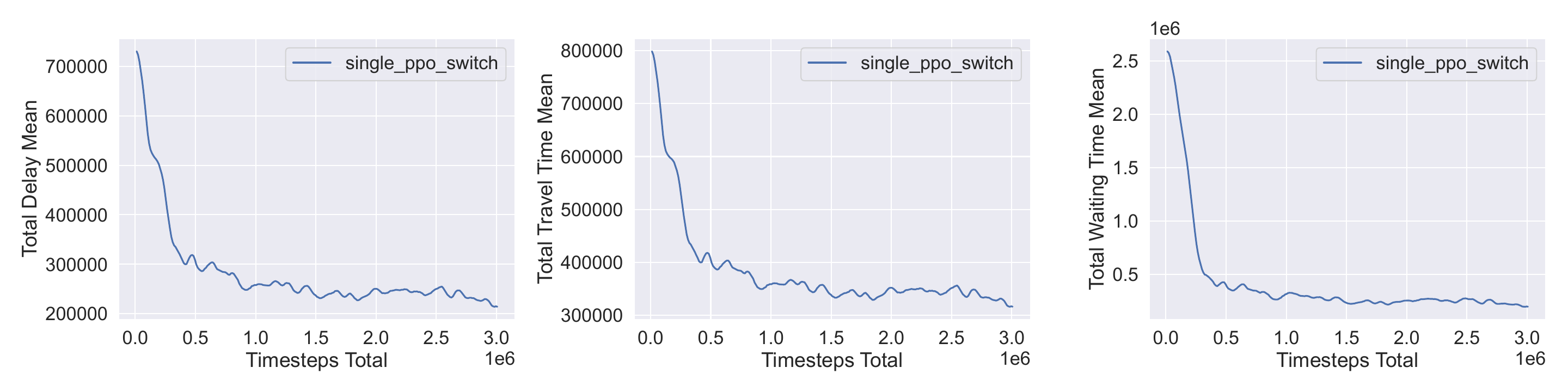}
        \caption{Single intersection}
        \label{fig:single_switch}
    \end{subfigure}

    \vspace{6pt}

    \begin{subfigure}[t]{1\linewidth}
        \centering
        \includegraphics[width=\linewidth]{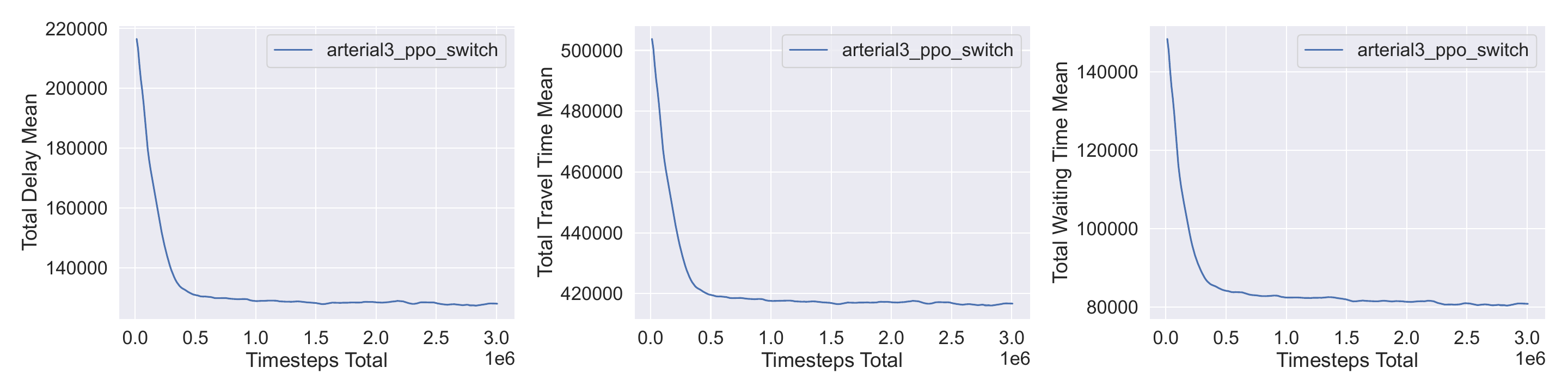}
        \caption{Arterial-3}
        \label{fig:arterial_switch}
    \end{subfigure}

    \caption{Training performance of {\small\texttt{SwitchNextOrNot}} in different scenarios (top: Single; bottom: Arterial-3).}
    \label{fig:results_switch}
\end{figure}

\begin{figure}[h]
    \centering
    \begin{subfigure}[t]{1\linewidth}
        \centering
        \includegraphics[width=\linewidth]{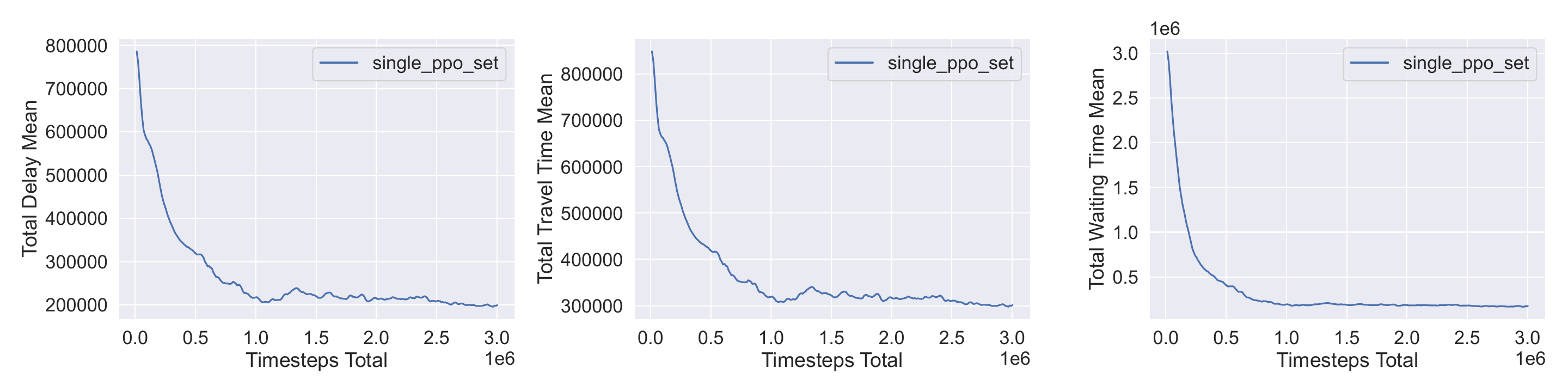}
        \caption{Single intersection}
        \label{fig:single_set}
    \end{subfigure}

    \vspace{6pt}

    \begin{subfigure}[t]{1\linewidth}
        \centering
        \includegraphics[width=\linewidth]{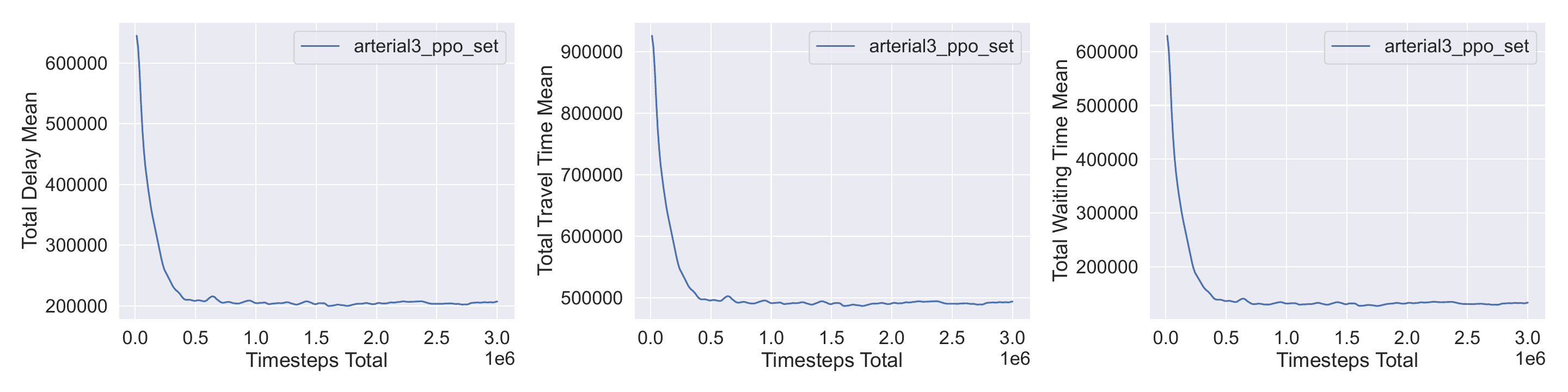}
        \caption{Arterial-3}
        \label{fig:arterial_set}
    \end{subfigure}

    \caption{Training performance of {\small\texttt{SetPhaseDuration}} in different scenarios (top: Single; bottom: Arterial-3).}
    \label{fig:results_set}
\end{figure}

The experimental results (Figures \ref{fig:results_choose} to \ref{fig:results_set}) show that, regardless of the action design, all three metrics decrease rapidly in the early stages of training, particularly within the first several hundred thousand steps, reflecting the agent's ability to quickly learn and adapt its signal control strategies. 
As training progresses, the rate of improvement gradually slows and eventually converges, indicating stable control performance. 
Comparing across scenarios, all three action designs exhibit consistent convergence in both the Single-Intersection case (Figures \ref{fig:single_choose}, \ref{fig:single_switch}, \ref{fig:single_set}) and the Arterial-3 case (Figures \ref{fig:arterial_choose}, \ref{fig:arterial_switch}, \ref{fig:arterial_set}), demonstrating that the proposed framework effectively enhances signal control in both isolated and coordinated multi-intersection settings. 
Overall, these results confirm the applicability and scalability of the VissimRL framework in diverse traffic scenarios.

\subsubsection{Case Study: Green Wave Emergence}
In the multi-agent experiments conducted on the three-intersection arterial network (Arterial-3), a clear Green Wave phenomenon was observed. 
A Green Wave refers to the coordinated timing of traffic signals that enables vehicles to pass through multiple intersections without stopping, thereby improving arterial efficiency and reducing delays. 
In this case study, the {\small\texttt{SwitchNextOrNot}} action design was applied, allowing each agent to decide at every decision step whether to maintain or switch the current phase, thus facilitating potential signal coordination.

\begin{figure}[h]
    \centering
    \includegraphics[width=0.9\linewidth]{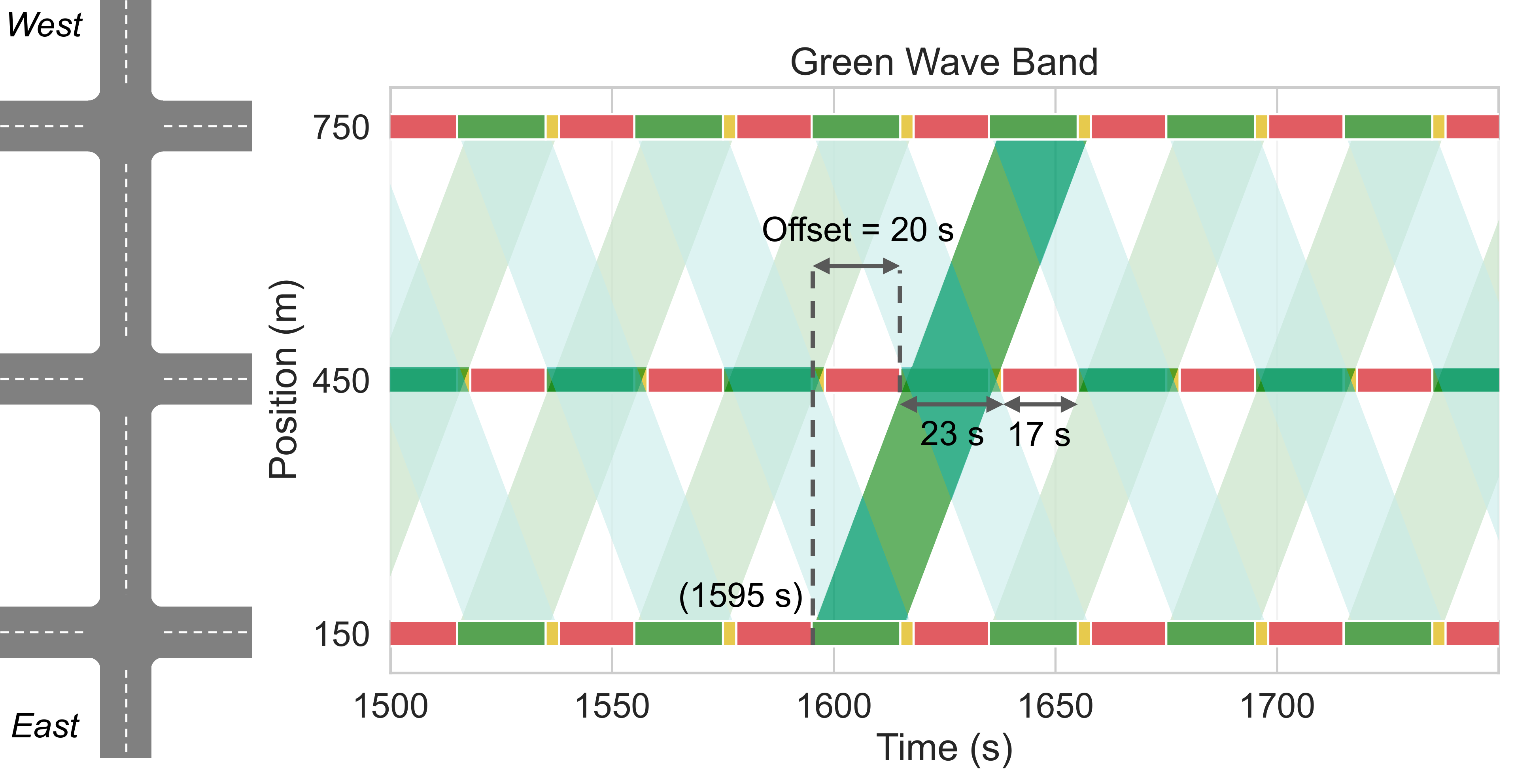}
    \caption{Green Wave Band visualization for the Arterial-3 case.}
    \label{fig:green_wave_band}
\end{figure}

To analyze this phenomenon, the results were visualized using a Green Wave Band diagram (Figure \ref{fig:green_wave_band}). 
In this diagram, the vertical axis represents the spatial position along the arterial, while the horizontal axis denotes time. 
The horizontal bands correspond to signal states at each intersection in the arterial direction, where green and red sections indicate green and red phases, respectively. 
Given the free-flow speed of 50 km/h, the slope of the green bands represents the distance–time ratio, reflecting the spatiotemporal trajectory of vehicles traveling at free-flow speed. 
When green phases appear continuous and aligned along this slope, it indicates effective signal coordination, enabling vehicles to travel along the arterial without stopping.

In this case, the arterial demand was set to 1200 vehicles per hour and the side-street demand to 800 vehicles per hour, with an intersection spacing of 300 meters and no turning movements (Table \ref{tab:synthetic_traffic_config}). 
Under these conditions, the optimal green split corresponds to the flow ratio of $1200:800\ (3:2)$, and the optimal offset is determined by the spacing–speed ratio, calculated as $300\ \text{m} / (50\ \text{km/h}) = 21.6\ \text{s}$. 
The experimental results (Figure \ref{fig:green_wave_band}) show that at simulation time 1595 seconds, the offset between the first and second eastbound intersections was 20 seconds, with green and red phases lasting 23 seconds and 17 seconds, respectively—values closely matching the theoretical optimum. 
Furthermore, the Green Wave Band analysis reveals that green phases along the arterial were continuously aligned in both directions on the time–space diagram, indicating that the agents successfully learned near-optimal coordination strategies. 
These results not only validate the effectiveness of the proposed VissimRL framework in multi-agent TSC but also demonstrate the potential of RL for coordinated signal optimization.

\subsubsection{Case Study: Dayuan Interchange}
\label{subsec:dayuan}
In the real-world case of the Dayuan Interchange, this study compares fixed-time control with RL using the PPO algorithm under peak-hour traffic conditions. 
Table \ref{tab:dayuan_performance_switch} reports the average results over five random seeds. 
PPO-based control shows improvements across all evaluation metrics. 
Under the {\small\texttt{SwitchNextOrNot}} action design, the average delay decreased from 559.712 seconds with fixed-time control to 208.962 seconds, representing a 63\% reduction. 
Average travel time and waiting time were also reduced by approximately 56\% and 38\%. 
Based on real traffic demand and a multi-intersection network, this case study highlights the potential of RL-TSC and further demonstrates the feasibility of applying the VissimRL framework in real-world traffic scenarios.
Additional performance comparisons with alternative action designs are provided in Appendix~\ref{appendix:action_comparison}.

\begin{table}[h]
    \centering
    \begin{threeparttable}
        \caption{Performance comparison at Dayuan Interchange.}
        \label{tab:dayuan_performance_switch}
        \begin{tabular}{lrrr}
            \toprule
            \textbf{TSC Method} & \textbf{Delay} & \textbf{Travel time} & \textbf{Waiting time} \\
            \midrule
            FixedTime & 559.712 & 634.960 & 802.311 \\
            \makecell[l]{PPO \\{\scriptsize(w/ \texttt{SwitchNextOrNot})}} & 208.962 & 281.028 & 494.420 \\
            \bottomrule
        \end{tabular}
        \begin{tablenotes}
            \footnotesize
            \item \textit{Note.} All performance metrics are lower-the-better and represent average per-vehicle values. 
            Evaluation was conducted with five different random seeds.
        \end{tablenotes}
    \end{threeparttable}
\end{table}

\section{Conclusions}
This paper introduced VissimRL, a modular and flexible RL framework for TSC integrated with the high-fidelity simulator Vissim. 
The framework effectively simplifies the development process by encapsulating the COM interface through a high-level Python API, while maintaining runtime efficiency and providing standardized environments for both single- and multi-agent learning. 
Particularly, the framework supports batching and caching of these APIs to improve the runtime efficiency while hiding the details of implementation. 
Experimental results demonstrate that VissimRL facilitates consistent improvements in traffic performance throughout the training process and enables emergent coordination among multiple agents. 
Its standardized architecture and integration with Vissim make it a practical bridge between academic RL research and industry-grade traffic simulation.
Future work will focus on extending the framework to more complex urban networks, exploring integration with existing traffic control strategies, and moving toward real-world deployment to further support the development of intelligent transportation systems.

\bibliographystyle{IEEEtran}
\bibliography{ref}

@misc{sumorl,
  title = {{{SUMO-RL}}},
  author = {Alegre, Lucas N.},
  year = {2019},
  publisher = {GitHub},
  journal = {GitHub repository},
  howpublished = {https://github.com/LucasAlegre/sumo-rl},
}

@article{baggioComparisonSimulatorsSUMO2023c,
  title = {Comparison of Simulators {{SUMO}} and {{VISSIM}} for Measuring Traffic Flow Interactions and Vehicle Emissions: A Case Study in {{Joinville}}, {{Santa Catarina}}, {{Brazil}}},
  shorttitle = {Comparison of Simulators {{SUMO}} and {{VISSIM}} for Measuring Traffic Flow Interactions and Vehicle Emissions},
  author = {Baggio, Mariana Luersen},
  year = {2023},
  publisher = {Joinville, SC.}
}

@misc{beedhamTrueEnvironmentalCost2022,
  title = {The {{True Environmental Cost}} of {{Inner-city Congestion}}},
  author = {Beedham, Matthew},
  year = {2022},
  journal = {TomTom},
  howpublished = {https://www.tomtom.com/newsroom/explainers-and-insights/the-true-environmental-cost-of-inner-city-congestion/},
  langid = {english}
}

@techreport{beestonTrafficModellingGuidelines2021,
  title = {Traffic {{Modelling Guidelines}} V4},
  author = {Beeston, Lucy and Blewitt, Robert and Bulmer, Sally and Wilson, James},
  year = {2021},
  month = sep,
  institution = {Transport for London},
  langid = {english}
}

@misc{daCityFlowEREfficientRealistic2024,
  title = {{{CityFlowER}}: {{An Efficient}} and {{Realistic Traffic Simulator}} with {{Embedded Machine Learning Models}}},
  shorttitle = {{{CityFlowER}}},
  author = {Da, Longchao and Chu, Chen and Zhang, Weinan and Wei, Hua},
  year = {2024},
  month = feb,
  number = {arXiv:2402.06127},
  eprint = {2402.06127},
  primaryclass = {cs},
  publisher = {arXiv},
  doi = {10.48550/arXiv.2402.06127},
  archiveprefix = {arXiv}
}

@misc{dlrandcontributorsSumolibPythonHelper,
  author = {{DLR and contributors}},
  title = {Sumolib: {{Python}} Helper Modules to Read Networks, Parse Output Data and Do Other Useful Stuff Related to the Traffic Simulation {{Eclipse SUMO}}},
  shorttitle = {Sumolib},
  howpublished = {https://sumo.dlr.de/docs/Tools/Sumolib.html}
}

@book{europeancourtofauditorsSustainableUrbanMobility2020a,
  title = {Sustainable Urban Mobility in the {{EU}}: No Substantial Improvement Is Possible without {{Member States}}' Commitment. {{Special}} Report {{No}} 06, 2020},
  shorttitle = {Sustainable Urban Mobility in the {{EU}}},
  author = {{European Court of Auditors}},
  year = {2020},
  publisher = {Publications Office of the European Union},
  doi = {10.2865/1094},
  isbn = {978-92-847-4341-4},
  langid = {english},
  lccn = {QJ-AB-20-003-EN-N}
}

@incollection{fellendorfMicroscopicTrafficFlow2010,
  title = {Microscopic {{Traffic Flow Simulator VISSIM}}},
  booktitle = {Fundamentals of {{Traffic Simulation}}},
  author = {Fellendorf, Martin and Vortisch, Peter},
  editor = {Barcel{\'o}, Jaume},
  year = {2010},
  pages = {63--93},
  publisher = {Springer},
  address = {New York, NY},
  doi = {10.1007/978-1-4419-6142-6_2},
  isbn = {978-1-4419-6142-6},
  langid = {english}
}

@techreport{hillDevelopmentMichiganSpecificVISSIM2020,
  title = {Development of a {{Michigan-Specific VISSIM Protocol}} for {{Submissions}} of {{VISSIM Modeling}}},
  author = {Hill, Matthew and Pittenger, Jason and Ceifetz, Andrew and WSP Michigan, {\relax Inc}.},
  year = {2020},
  month = aug,
  number = {SPR-1689},
  langid = {english},
  lccn = {dot:62525}
}

@techreport{hunterVISSIMSimulationGuidance2021,
  title = {{{VISSIM Simulation Guidance}}},
  author = {Hunter, Michael and {Georgia Institute of Technology}},
  year = {2021},
  month = jun,
  number = {FHWA-GA-21-1833, 18-33},
  langid = {english},
  lccn = {dot:60642}
}

@techreport{iowadepartmentoftransportationIowaDOTMicrosimulation2017,
  title = {Iowa {{DOT Microsimulation Guidance}}},
  author = {{Iowa Department of Transportation}},
  year = {2017},
  month = oct
}

@inproceedings{krajzewiczSUMOSimulationUrban2002,
  title = {{{SUMO}} ({{Simulation}} of {{Urban MObility}}) - an Open-Source Traffic Simulation},
  booktitle = {Proceedings of the 4th {{Middle East Symposium}} on {{Simulation}} and {{Modelling}} ({{MESM20002}})},
  author = {Krajzewicz, Daniel and Hertkorn, Georg and R{\"o}ssel, C. and Wagner, Peter},
  editor = {{Al-Akaidi}, A.},
  year = {2002},
  pages = {183--187},
  address = {Sharjah (United Arab Emirates)},
  isbn = {978-90-77039-09-0},
  langid = {english}
}

@misc{liangRLlibAbstractionsDistributed2017b,
  title = {{{RLlib}}: {{Abstractions}} for {{Distributed Reinforcement Learning}}},
  shorttitle = {{{RLlib}}},
  author = {Liang, Eric and Liaw, Richard and Moritz, Philipp and Nishihara, Robert and Fox, Roy and Goldberg, Ken and Gonzalez, Joseph E. and Jordan, Michael I. and Stoica, Ion},
  year = {2017},
  publisher = {arXiv},
  doi = {10.48550/ARXIV.1712.09381},
  copyright = {arXiv.org perpetual, non-exclusive license}
}

@article{michailidisTrafficSignalControl2025,
  title = {Traffic {{Signal Control}} via {{Reinforcement Learning}}: {{A Review}} on {{Applications}} and {{Innovations}}},
  shorttitle = {Traffic {{Signal Control}} via {{Reinforcement Learning}}},
  author = {Michailidis, Panagiotis and Michailidis, Iakovos and Lazaridis, Charalampos Rafail and Kosmatopoulos, Elias},
  year = {2025},
  month = may,
  journal = {Infrastructures},
  volume = {10},
  number = {5},
  pages = {114},
  publisher = {Multidisciplinary Digital Publishing Institute},
  issn = {2412-3811},
  doi = {10.3390/infrastructures10050114},
  copyright = {http://creativecommons.org/licenses/by/3.0/},
  langid = {english}
}

@misc{mnihPlayingAtariDeep2013,
  title = {Playing {{Atari}} with {{Deep Reinforcement Learning}}},
  author = {Mnih, Volodymyr and Kavukcuoglu, Koray and Silver, David and Graves, Alex and Antonoglou, Ioannis and Wierstra, Daan and Riedmiller, Martin},
  year = {2013},
  month = dec,
  number = {1312.5602},
  eprint = {1312.5602},
  primaryclass = {cs},
  publisher = {arXiv},
  doi = {10.48550/arXiv.1312.5602},
  archiveprefix = {arXiv}
}

@article{noaeenReinforcementLearningUrban2022,
  title = {Reinforcement Learning in Urban Network Traffic Signal Control: {{A}} Systematic Literature Review},
  shorttitle = {Reinforcement Learning in Urban Network Traffic Signal Control},
  author = {Noaeen, Mohammad and Naik, Atharva and Goodman, Liana and Crebo, Jared and Abrar, Taimoor and Abad, Zahra Shakeri Hossein and Bazzan, Ana L. C. and Far, Behrouz},
  year = {2022},
  month = aug,
  journal = {Expert Systems with Applications},
  volume = {199},
  pages = {116830},
  issn = {0957-4174},
  doi = {10.1016/j.eswa.2022.116830}
}

@inproceedings{oroojlooyAttendLightUniversalAttentionBased2020,
  title = {{{AttendLight}}: {{Universal Attention-Based Reinforcement Learning Model}} for {{Traffic Signal Control}}},
  shorttitle = {{{AttendLight}}},
  booktitle = {Advances in {{Neural Information Processing Systems}}},
  author = {Oroojlooy, Afshin and Nazari, Mohammadreza and Hajinezhad, Davood and Silva, Jorge},
  year = {2020},
  volume = {33},
  pages = {4079--4090},
  publisher = {Curran Associates, Inc.}
}

@article{raffinStablebaselines3ReliableReinforcement2021,
  title = {Stable-Baselines3: Reliable Reinforcement Learning Implementations},
  shorttitle = {Stable-Baselines3},
  author = {Raffin, Antonin and Hill, Ashley and Gleave, Adam and Kanervisto, Anssi and Ernestus, Maximilian and Dormann, Noah},
  year = {2021},
  month = jan,
  journal = {J. Mach. Learn. Res.},
  volume = {22},
  number = {1},
  pages = {268:12348--268:12355},
  issn = {1532-4435}
}

@techreport{schilperoortProtocolVISSIMSimulation2014,
  title = {Protocol for {{VISSIM Simulation}}},
  author = {Schilperoort, LisaRene and McClanahan, Doug and Shank, Ray and Bjordahl, Mike},
  year = {2014},
  month = sep,
  institution = {Washington State Department of Transportation},
  langid = {english}
}

@article{schraderExtensionValidationNEMAStyle2022,
  title = {Extension and {{Validation}} of {{NEMA-Style Dual-Ring Controller}} in {{SUMO}}},
  author = {Schrader, Max and Wang, Qichao and Bittle, Joshua},
  year = {2022},
  month = sep,
  journal = {SUMO Conference Proceedings},
  volume = {3},
  pages = {1--13},
  issn = {2750-4425},
  doi = {10.52825/scp.v3i.115},
  copyright = {Copyright (c) 2022 Max Schrader, Qichao Wang, Joshua Bittle},
  langid = {english}
}

@article{schrank2023UrbanMobility2024,
  title = {2023 {{Urban Mobility Report}}},
  author = {Schrank, David and Albert, Luke and Jha, Kartikeya and Eisele, Bill},
  year = {2024},
  month = jun,
  langid = {american}
}

@misc{schulmanProximalPolicyOptimization2017,
  title = {Proximal {{Policy Optimization Algorithms}}},
  author = {Schulman, John and Wolski, Filip and Dhariwal, Prafulla and Radford, Alec and Klimov, Oleg},
  year = {2017},
  month = aug,
  number = {arXiv:1707.06347},
  eprint = {1707.06347},
  primaryclass = {cs},
  publisher = {arXiv},
  doi = {10.48550/arXiv.1707.06347},
  archiveprefix = {arXiv}
}

@inproceedings{shashiStudyDeepReinforcement2021,
  title = {A {{Study}} on {{Deep Reinforcement Learning Based Traffic Signal Control}} for {{Mitigating Traffic Congestion}}},
  booktitle = {2021 {{IEEE}} 3rd {{Eurasia Conference}} on {{Biomedical Engineering}}, {{Healthcare}} and {{Sustainability}} ({{ECBIOS}})},
  author = {Shashi, Farjana Islam and Md Sultan, Salman and Khatun, Afroza and Sultana, Tangina and Alam, Tahira},
  year = {2021},
  month = may,
  pages = {288--291},
  doi = {10.1109/ECBIOS51820.2021.9510422}
}

@misc{whiteCOMObjectsInterfaces,
  title = {{{COM Objects}} and {{Interfaces}} - {{Win32}} Apps},
  author = {White, Steven},
  howpublished = {https://learn.microsoft.com/en-us/windows/win32/com/com-objects-and-interfaces},
  langid = {american}
}

@misc{terryPettingZooGymMultiAgent2021,
  title = {{{PettingZoo}}: {{Gym}} for {{Multi-Agent Reinforcement Learning}}},
  shorttitle = {{{PettingZoo}}},
  author = {Terry, J. K. and Black, Benjamin and Grammel, Nathaniel and Jayakumar, Mario and Hari, Ananth and Sullivan, Ryan and Santos, Luis and Perez, Rodrigo and Horsch, Caroline and Dieffendahl, Clemens and Williams, Niall L. and Lokesh, Yashas and Ravi, Praveen},
  year = {2021},
  month = oct,
  number = {arXiv:2009.14471},
  eprint = {2009.14471},
  primaryclass = {cs},
  publisher = {arXiv},
  doi = {10.48550/arXiv.2009.14471},
  archiveprefix = {arXiv}
}

@misc{towersGymnasiumStandardInterface2024,
  title = {Gymnasium: {{A Standard Interface}} for {{Reinforcement Learning Environments}}},
  shorttitle = {Gymnasium},
  author = {Towers, Mark and Kwiatkowski, Ariel and Terry, Jordan and Balis, John U. and Cola, Gianluca De and Deleu, Tristan and Goul{\~a}o, Manuel and Kallinteris, Andreas and Krimmel, Markus and KG, Arjun and {Perez-Vicente}, Rodrigo and Pierr{\'e}, Andrea and Schulhoff, Sander and Tai, Jun Jet and Tan, Hannah and Younis, Omar G.},
  year = {2024},
  month = nov,
  number = {arXiv:2407.17032},
  eprint = {2407.17032},
  primaryclass = {cs},
  publisher = {arXiv},
  doi = {10.48550/arXiv.2407.17032},
  archiveprefix = {arXiv}
}

@article{vanderpolCoordinatedDeepReinforcement2016,
  title = {Coordinated Deep Reinforcement Learners for Traffic Light Control},
  author = {{Van der Pol}, Elise and Oliehoek, Frans A.},
  year = {2016},
  journal = {Proceedings of learning, inference and control of multi-agent systems (at NIPS 2016)},
  volume = {8},
  pages = {21--38}
}

@article{wangTrafficSignalCycle2024,
  title = {Traffic {{Signal Cycle Control With Centralized Critic}} and {{Decentralized Actors Under Varying Intervention Frequencies}}},
  author = {Wang, Maonan and Chen, Yirong and Kan, Yuheng and Xu, Chengcheng and Lepech, Michael and Pun, Man-On and Xiong, Xi},
  year = {2024},
  month = feb,
  journal = {IEEE Transactions on Intelligent Transportation Systems},
  volume = {25},
  number = {12},
  pages = {20085--20104},
  issn = {1558-0016},
  doi = {10.1109/TITS.2024.3462153}
}

@inproceedings{weiCoLightLearningNetworklevel2019,
  title = {{{CoLight}}: {{Learning Network-level Cooperation}} for {{Traffic Signal Control}}},
  shorttitle = {{{CoLight}}},
  booktitle = {Proceedings of the 28th {{ACM International Conference}} on {{Information}} and {{Knowledge Management}}},
  author = {Wei, Hua and Xu, Nan and Zhang, Huichu and Zheng, Guanjie and Zang, Xinshi and Chen, Chacha and Zhang, Weinan and Zhu, Yanmin and Xu, Kai and Li, Zhenhui},
  year = {2019},
  month = nov,
  series = {{{CIKM}} '19},
  pages = {1913--1922},
  publisher = {Association for Computing Machinery},
  address = {New York, NY, USA},
  doi = {10.1145/3357384.3357902},
  isbn = {978-1-4503-6976-3}
}

@inproceedings{weiIntelliLightReinforcementLearning2018,
  title = {{{IntelliLight}}: {{A Reinforcement Learning Approach}} for {{Intelligent Traffic Light Control}}},
  shorttitle = {{{IntelliLight}}},
  booktitle = {Proceedings of the 24th {{ACM SIGKDD International Conference}} on {{Knowledge Discovery}} \& {{Data Mining}}},
  author = {Wei, Hua and Zheng, Guanjie and Yao, Huaxiu and Li, Zhenhui},
  year = {2018},
  month = jul,
  series = {{{KDD}} '18},
  pages = {2496--2505},
  publisher = {Association for Computing Machinery},
  address = {New York, NY, USA},
  doi = {10.1145/3219819.3220096},
  isbn = {978-1-4503-5552-0}
}

@inproceedings{weiPressLightLearningMax2019,
  title = {{{PressLight}}: {{Learning Max Pressure Control}} to {{Coordinate Traffic Signals}} in {{Arterial Network}}},
  shorttitle = {{{PressLight}}},
  booktitle = {Proceedings of the 25th {{ACM SIGKDD International Conference}} on {{Knowledge Discovery}} \& {{Data Mining}}},
  author = {Wei, Hua and Chen, Chacha and Zheng, Guanjie and Wu, Kan and Gayah, Vikash and Xu, Kai and Li, Zhenhui},
  year = {2019},
  month = jul,
  series = {{{KDD}} '19},
  pages = {1290--1298},
  publisher = {Association for Computing Machinery},
  address = {New York, NY, USA},
  doi = {10.1145/3292500.3330949},
  isbn = {978-1-4503-6201-6}
}

@misc{yuSurprisingEffectivenessPPO2022,
  title = {The {{Surprising Effectiveness}} of {{PPO}} in {{Cooperative}}, {{Multi-Agent Games}}},
  author = {Yu, Chao and Velu, Akash and Vinitsky, Eugene and Gao, Jiaxuan and Wang, Yu and Bayen, Alexandre and Wu, Yi},
  year = {2022},
  month = nov,
  number = {arXiv:2103.01955},
  eprint = {2103.01955},
  primaryclass = {cs},
  publisher = {arXiv},
  doi = {10.48550/arXiv.2103.01955},
  archiveprefix = {arXiv}
}

@inproceedings{zhangCityFlowMultiAgentReinforcement2019,
  title = {{{CityFlow}}: {{A Multi-Agent Reinforcement Learning Environment}} for {{Large Scale City Traffic Scenario}}},
  shorttitle = {{{CityFlow}}},
  booktitle = {The {{World Wide Web Conference}}},
  author = {Zhang, Huichu and Feng, Siyuan and Liu, Chang and Ding, Yaoyao and Zhu, Yichen and Zhou, Zihan and Zhang, Weinan and Yu, Yong and Jin, Haiming and Li, Zhenhui},
  year = {2019},
  month = may,
  series = {{{WWW}} '19},
  pages = {3620--3624},
  publisher = {Association for Computing Machinery},
  address = {New York, NY, USA},
  doi = {10.1145/3308558.3314139},
  isbn = {978-1-4503-6674-8}
}

@article{zhaoSurveyDeepReinforcement2024,
  title = {A Survey on Deep Reinforcement Learning Approaches for Traffic Signal Control},
  author = {Zhao, Haiyan and Dong, Chengcheng and Cao, Jian and Chen, Qingkui},
  year = {2024},
  month = jul,
  journal = {Engineering Applications of Artificial Intelligence},
  volume = {133},
  pages = {108100},
  issn = {0952-1976},
  doi = {10.1016/j.engappai.2024.108100}
}

@misc{zhengDiagnosingReinforcementLearning2019,
  title = {Diagnosing {{Reinforcement Learning}} for {{Traffic Signal Control}}},
  author = {Zheng, Guanjie and Zang, Xinshi and Xu, Nan and Wei, Hua and Yu, Zhengyao and Gayah, Vikash and Xu, Kai and Li, Zhenhui},
  year = {2019},
  month = may,
  number = {arXiv:1905.04716},
  eprint = {1905.04716},
  primaryclass = {cs},
  publisher = {arXiv},
  doi = {10.48550/arXiv.1905.04716},
  archiveprefix = {arXiv}
}

@inproceedings{Chang_VissimRL_IV,
  title     = {VissimRL: A Multi-Agent Reinforcement Learning Framework for Traffic Signal Control Based on Vissim},
  author    = {Chang, Hsiao-Chuan and
               Huang, Sheng-You and
               Chen, Yen-Chi and
               Wu, I-Chen},
  booktitle = {2026 IEEE Intelligent Vehicles Symposium (IV)},
  year      = {2026},
  publisher = {IEEE}
}

\clearpage
\onecolumn
\appendices
\section{Configuration Parameters of the VissimRL Framework}
This appendix presents the detailed configuration parameters of the proposed VissimRL framework. 
The framework consists of two main components: the Vissim Wrapper, which interfaces with the traffic simulator Vissim, and the RL Environment, which provides a flexible and customizable interface for agent-environment interaction.

\renewcommand{\thesubsection}{\Alph{section}-\arabic{subsection}}
\renewcommand{\thesubsectiondis}{\thesubsection}
\setcounter{subsection}{0}
\subsection{Vissim Wrapper Configuration}
\label{appendix:wrapper_config}
The VissimWrapper serves as the core interface for simulation. 
It allows users to initialize and control the Vissim simulation environment through Python, offering a set of configurable parameters that define the network files and simulation settings.  

Table~\ref{tab:vissim_config} lists all supported configuration options. 
Parameters such as {\small\texttt{net\_path}}, {\small\texttt{sig\_path}}, and {\small\texttt{detector\_path}} must be provided by the user, as they specify the essential input files for network structure, signal control, and detector placement. 
Other parameters such as {\small\texttt{use\_gui}}, {\small\texttt{quick\_mode}}, and {\small\texttt{sim\_res}} control simulation performance and visualization options. 

\begin{table}[H]
    \centering
    \begin{threeparttable}
    \caption{Configuration parameters of the Vissim Wrapper.}
    \label{tab:vissim_config}
    \begin{tabular}{>{\ttfamily}l >{\ttfamily}l >{\ttfamily}l l}
        \toprule
        {\normalfont\bfseries Parameter} & {\normalfont\bfseries Type} &
        {\normalfont\bfseries Default} & {\normalfont\bfseries Description} \\ \midrule
        net\_path & str  & --   & Path to the Vissim network file \\
        sig\_path & str  & --   & Path to the signal control file \\
        detector\_path & str  & --   & Path to the detector configuration file \\
        use\_gui & bool & False & Whether to run simulation with GUI \\
        quick\_mode & bool & True  & Enable faster simulation mode \\
        start\_time & int  & 600  & Simulation start time (s) \\
        sim\_period & int  & 4200 & Total simulation duration (s) \\
        sim\_res & int  & 10   & Simulation resolution (time steps per second) \\
        seed & int  & 42   & Random seed for reproducibility \\ \bottomrule
    \end{tabular}
    \begin{tablenotes}
        \footnotesize
        \item \textit{Note.} Use \texttt{vissim = VissimWrapper(**config)} to create a simulation environment.
        \item Parameters marked with “--” do not have default values and must be specified by the user.
    \end{tablenotes}
    \end{threeparttable}
\end{table}

\subsection{Reinforcement Learning Environment Configuration}
\label{appendix:rlenv_config}
The RL Environment Framework provides standardized interfaces for reinforcement learning, built upon the Gymnasium and PettingZoo APIs to support both single-agent and multi-agent training.
It offers configurable settings for simulation control, signal phase constraints, reinforcement learning behavior, and logging management.

Users can instantiate the environment with a single line of code using {\small\texttt{gym.make(**config)}} for single-agent training or {\small\texttt{parallel\_env(**config)}} for multi-agent training. 
Table~\ref{tab:rl_env_config} summarizes all configuration parameters, categorized into four groups: simulation settings, traffic signal settings, RL settings, and logger settings. The simulation settings are identical to those of the Vissim Wrapper, as the RL environment automatically initializes the wrapper during instantiate.

\begin{ThreePartTable}
    \begin{TableNotes}
        \footnotesize
        \item \textit{Note.} The environment can be instantiated with a single line: 
        \item \hspace*{2.5em}- \texttt{gym.make(**config)} for single-agent
        \item \hspace*{2.5em}- \texttt{parallel\_env(**config)} for multi-agent training.
        \item \hspace*{2.5em}Parameters marked with “--” have no default values and must be specified by the user.
    \end{TableNotes}

    {\footnotesize
    \begin{longtable}{>{\ttfamily}l >{\ttfamily}l >{\ttfamily}l p{5.5cm}}
        \caption{RL Environment Framework configuration parameters.}
        \label{tab:rl_env_config} \\
        \toprule
        {\normalfont\bfseries Parameter} & {\normalfont\bfseries Type} & {\normalfont\bfseries Default Setting} & {\normalfont\bfseries Description} \\ \midrule
        \endfirsthead
    
        \multicolumn{4}{c}{\textbf{Table \thetable\ (continued)}} \\ \toprule
        {\normalfont\bfseries Parameter} & {\normalfont\bfseries Type} & {\normalfont\bfseries Default Setting} & {\normalfont\bfseries Description} \\ \midrule
        \endhead
    
        \midrule
        \multicolumn{4}{r}{\textit{Continued on next page}} \\
        \endfoot
    
        \bottomrule
        \insertTableNotes
        \endlastfoot
    
        \multicolumn{4}{l}{\textit{Simulation settings}} \\
        net\_path & str & -- & File path of the Vissim network model \\
        sig\_path & str & -- & File path of the signal control configuration \\
        detector\_path & str & -- & File path of the detector configuration \\
        use\_gui & bool & False & Run simulation with graphical user interface \\
        quick\_mode & bool & True & Enable faster simulation mode \\
        start\_time & int & 600 & Simulation start time (s) \\
        sim\_period & int & 4200 & Total simulation duration (s) \\
        sim\_res & int & 10 & Simulation resolution (steps per second) \\
        seed & int & 42 & Random seed for reproducibility \\ \midrule
    
        \multicolumn{4}{l}{\textit{Traffic signal settings}} \\
        yellow\_time & int & 3 & Duration of yellow phase (s) \\
        intergreen\_time & int & 5 & Duration of the intergreen phase (s) \\
        min\_green & int & 8 & Minimum green phase duration (s) \\
        max\_green & int & 120 & Maximum green phase duration (s) \\ \pagebreak
    
        \multicolumn{4}{l}{\textit{RL settings}} \\
        single\_agent & bool & False & Whether to returns single values like regular Gym style \\
        ctrl\_intersections & list/str & "all" & Controlled intersections (list of IDs or 'all') \\
        observation\_config & dict & \makecell[tl]{\{class:\\\hspace{1em}DefaultObservationFunction,\\\ args: \{\}\}} & Observation function and arguments \\
        action\_config & dict & \makecell[tl]{\{class:\\\hspace{1em}SwitchNextOrNot,\\\ args: \{\}\}} & Action function and arguments \\
        reward\_config & dict & \makecell[tl]{\{class:\\\hspace{1em}DefaultRewardFunction,\\\ args: \{\}\}} & Reward function and arguments \\ \midrule
    
        \multicolumn{4}{l}{\textit{Logger settings}} \\
        enable\_loggers & None/list/str & ["vissim\_rl.environment.env"] & Enabled loggers ('all', None, or list of module paths) \\
        log\_dir & str & "logs" & Directory for log files \\
        log\_suffix & str & "VissimRL" & Log file name suffix \\
        log\_to\_file & bool & False & Whether to write logs to file \\
        log\_to\_console & bool & True & Whether to output logs to console \\
        log\_level & str & "INFO" & Logging level (e.g., DEBUG, INFO, ERROR) \\
    \end{longtable}
    }
\end{ThreePartTable}

\section{RL Settings for Performance Validation}
\label{appendix:rl_settings}
Table~\ref{tab:rl_settings} summarizes the RL configurations used in the performance validation experiments of the proposed VissimRL framework. 
These settings were used to train the RL agents under consistent conditions across both single-agent and multi-agent scenarios.
The Proximal Policy Optimization (PPO) algorithm implemented in RLlib was adopted for all experiments. 
Different observation functions were used to accommodate the scope of perception in each scenario, while the reward function followed a unified multi-objective design that combines internal and boundary waiting times, throughput, and average speed with their respective weights. 
This configuration ensured comparability of performance metrics across experimental settings.

\begin{table}[h]
    \centering
    \begin{threeparttable}
    \caption{RL settings for performance validation experiments.}
    \label{tab:rl_settings}
    \begin{tabular}{ll}
        \toprule
        \textbf{Item} & \textbf{Setting} \\
        \midrule
        RL algorithm & PPO \\
        Observation & \makecell[lt]{Single-agent: \texttt{LocalObservationFunction}\\
        Multi-agent: \texttt{GlobalObservationFunction}}\\
        Reward function & \texttt{DefaultRewardFunction} (multi-objective weights) \\ 
        \addlinespace[3pt] 
        \hdashline[3pt/1.5pt]
        \addlinespace[3pt] 
        \quad\textit{- itwt (Internal waiting time)} & \(-0.0001\) \\
        \quad\textit{- btwt (Boundary waiting time)} & \(-0.0001\) \\
        \quad\textit{- throughput} & \(0.01\) \\
        \quad\textit{- speed} & \(0.001\) \\ \bottomrule
    \end{tabular}
    \begin{tablenotes}
        \footnotesize
        \item \textit{Note.} Observation functions differ by scenario (single vs. multi-agent); Reward function applies the same weighted combination across all experiments.
    \end{tablenotes}
    \end{threeparttable}
\end{table}

\section{Supplementary Performance Comparison at Dayuan Interchange}
\label{appendix:action_comparison}

Table~\ref{tab:action_comparison} presents a supplementary comparison of three action designs implemented within the proposed VissimRL framework. 
While the main text (Section~\ref{subsec:dayuan}) discusses the {\small\texttt{SwitchNextOrNot}} action as the primary design due to its stable and well-coordinated control performance, two additional action designs—{\small\texttt{ChooseNextPhase}} and {\small\texttt{SetPhaseDuration}}—were evaluated to provide a more comprehensive understanding of the framework's behavior under different control paradigms. 
All experiments were conducted at the Dayuan Interchange using identical network and training configurations, and the reported results are averaged over five random seeds for statistical consistency.  

The {\small\texttt{SwitchNextOrNot}} action serves as the baseline reinforcement learning design, where agents decide whether to switch to the next predefined phase or keep the current one. 
This constraint simplifies the coordination process among intersections and yields stable performance across all evaluated metrics.  

The {\small\texttt{ChooseNextPhase}} action extends flexibility by allowing agents to directly select any subsequent phase, unconstrained by a fixed sequence. 
This design improves responsiveness and adaptability to local traffic variations, resulting in lower average delay and travel time. 
However, because agents make independent phase-selection decisions, coordination across intersections becomes more challenging compared to the {\small\texttt{SwitchNextOrNot}} design, leading to congestion at the network boundaries and less pronounced improvement in average travel time.  

The {\small\texttt{SetPhaseDuration}} action enables agents to determine the exact duration for each phase. 
This design reduces the frequency of decisions and enables agents to perform more detailed control. However, asynchronous phase adjustments across intersections still hinder network-level coordination, ultimately leading to degraded overall performance.
Future work could explore incorporating temporal models such as recurrent neural networks (RNNs) or multi-agent learning schemes like centralized training with decentralized execution (CTDE) to improve coordination and temporal stability across agents.  


\begin{table}[h]
    \centering
    \begin{threeparttable}
        \caption{Performance comparison among three action designs at Dayuan Interchange.}
        \label{tab:action_comparison}
        \begin{tabular}{lrrr}
            \toprule
            \textbf{TSC Method} & \textbf{Delay} & \textbf{Travel time} & \textbf{Waiting time} \\
            \midrule
            FixedTime & 559.712 & 634.960 & 802.311 \\
            PPO (w/ \texttt{SwitchNextOrNot} action) & 208.962 & 281.028 & 494.420 \\
            PPO (w/ \texttt{ChooseNextPhase} action) & 157.016 & 227.907 & 543.741 \\
            PPO (w/ \texttt{SetPhaseDuration} action) & 842.715 & 910.755 & 1,234.939 \\
            \bottomrule
        \end{tabular}
        \begin{tablenotes}
            \footnotesize
            \item \textit{Note.} All performance metrics are lower-the-better and represent average per-vehicle values. 
            Evaluation was conducted with five different random seeds.
        \end{tablenotes}
    \end{threeparttable}
\end{table}

\end{document}